\documentclass[fleqn,usenatbib]{mnras}

\usepackage{orcidlink}
\usepackage{graphicx}
\usepackage{amsmath}
\usepackage{amssymb}
\usepackage{multicol}
\usepackage{bm}
\usepackage{pdflscape}
\usepackage{amsfonts}
\usepackage{txfonts}
\usepackage{ctable}
\usepackage{url}
\usepackage{lineno}
\usepackage[normalem]{ulem}
\usepackage{color}
\usepackage{grffile}
\usepackage{booktabs}
\usepackage[american]{babel}

\newcommand\altaffilmark[1]{$^{#1}$}
\newcommand\altaffiltext[1]{$^{#1}$}
\newcommand{\fraction}[2]{\dfrac{\displaystyle#1}{\displaystyle#2}}
\newcommand{\pder}[2]{\fraction{\partial#1}{\partial#2}}
\newcommand{\pdder}[2]{\fraction{\partial^2#1}{\partial#2^2}}
\newcommand{\gvel}{\mathbf{u}}
\newcommand{\pR}{p_{\rm R}}
\newcommand{\pL}{p_{\rm L}}
\newcommand{\SLconst}{C_{\rm s}}

\usepackage[T1]{fontenc}
\usepackage{ae,aecompl}

\definecolor{linkcolor}{rgb}{0.0,0.3,0.5}
\usepackage{hyperref}
\hypersetup{
    colorlinks=true,
    linkcolor=linkcolor,
    citecolor=linkcolor,
    filecolor=linkcolor,
    urlcolor=linkcolor,
}
\graphicspath{{./figures/}}

\title[It\^o tracers]{It\^o tracers: continuous-trajectory Lagrangian particles for Eulerian hydrodynamics}

\author[Moseley, Teyssier, \& Abel]{
\parbox[t]{\textwidth}{
    Eric R.~Moseley\altaffilmark{1,2,3}\thanks{E-mail: emoseley@stanford.edu}\orcidlink{0000-0001-8558-5009},
    R.~Teyssier\altaffilmark{4}\orcidlink{0000-0001-7689-0933},
    Tom Abel\altaffilmark{1,2,3}\orcidlink{0000-0002-5969-1251}
}
\vspace*{6pt} \\
\altaffiltext{1}{Kavli Institute for Particle Astrophysics and Cosmology, Stanford University, 452 Lomita Mall, Stanford, CA 94305, USA}\\
\altaffiltext{2}{Department of Physics, Stanford University, 382 Via Pueblo Mall, Stanford, CA 94305, USA}\\
\altaffiltext{3}{SLAC National Accelerator Laboratory, 2575 Sand Hill Road, Menlo Park, CA 94025, USA}\\
\altaffiltext{4}{Department of Astrophysical Sciences, Princeton University, Princeton, NJ 08540, USA}\\
}

\date{}
\pubyear{2026}

\begin{document}
\maketitle

\begin{abstract}
Lagrangian tracer particles have long been used to track the history of individual gas parcels in hydrodynamical codes. Particles advected by the cell-centered velocity carry no representation of underlying numerical diffusion, and thus exhibit systematic bias. The Monte-Carlo (MC) tracer resolves this with discrete probabilistic cell-to-cell, flux-based jumps, at the cost of trajectories that are discontinuous in time. We introduce the It\^o tracer, a continuous-time Lagrangian particle with moments matched to the advection, diffusion, and dispersion of the gas. A subgrid-scale variant (SGS-It\^o) replaces the numerical diffusion with a Smagorinsky--Lilly turbulent diffusivity, illustrating that the form of the diffusion matters less than its magnitude. We validate these methods with a 1D square-pulse advection test and 3D decaying turbulence at $\sigma_{\rm rms} = 15\,c_{\rm s}$.
We compare the different tracer particle methods using several statistical tests. It\^o tracers largely reproduce or improve upon MC tracers statistics across column-density maps, joint density histograms, log-density-ratio PDFs, and density power spectra. In the turbulence test, It\^o tracers improve the correlation between tracers and gas over the MC tracers by >3\%, and reduce the width of the log-density ratio PDF by nearly 50\%. Relative to classical tracers, these improvements are $\gtrsim$30\% and 230\%, respectively. Because It\^o tracers follow a stochastic differential equation, the method maps onto other continuous-trajectory Lagrangian processes (e.g. dust grains, charged particles, cosmic rays), admits variance-reduction techniques, higher-order integrators, and GPU-friendly implementations -- all of which are unavailable to discrete-jump schemes.
\end{abstract}


\section{Introduction}
\label{sec:intro}

Many astrophysical problems require knowing the history of individual fluid parcels. Finite-volume codes such as \textsc{RAMSES} \citep{teyssier2002cosmological}, \textsc{ATHENA++} \citep{stone2020athena}, \textsc{ENZO} \citep{oshea2005enzo}, and the moving-mesh code \textsc{AREPO} \citep{springel2010pur} discretize the fluid into cells between which fluxes are passed, and so do not carry this information by default: They naturally mix and diffuse the fluid. The continuum trajectories of fluid parcels are lost at each time-step as an unavoidable consequence of numerical diffusion. Smoothed-particle hydrodynamics \citep[SPH;][]{price2018phantom} and meshless finite-mass methods \citep[MFM;][]{Hopkins_2015} allow for built-in history of fluid parcels, because each particle is itself a Lagrangian mass element. They pay for this with spatially inhomogeneous resolution and, in some regimes, with reduced accuracy at shocks and shear layers \citep{agertz2007fundamental,springel2010smoothed}. As well, because they do not solve the mass conservation equation explicitly, such methods tend to under-mix chemical species; restoring this property requires solving a diffusion equation sampled at the particle locations \citep{tasker2008test,su2017feedback}. The perfect Lagrangian fidelity of such methods is thus both an advantage and a disadvantage.

Further, astrophysical plasmas frequently have a dynamically non-negligible magnetic field, with Alfv\'en Mach number $M_{\rm A} \lesssim 1$ and plasma-$\beta$ $\lesssim 1$ \citep[e.g., in molecular clouds;][]{crutcher2012magnetic}. In these regimes, finite-volume, Eulerian codes have a clear advantage over Lagrangian codes, in that they may use the constrained transport algorithm, enabling the preservation of $\nabla\cdot{\mathbf B} = 0$ to machine precision \citep{evans1988simulation,fromang2006high}.\footnote{A possible exception is \textsc{AREPO}, for which a CT algorithm has been implemented \citep{Mocz2014}. However, \textsc{AREPO} is not, strictly speaking, a Lagrangian code, but something of a hybrid scheme.} As Lagrangian schemes have no natural staggered structure, they must rely instead on divergence-cleaning methods that either dissipate magnetic energy or complicate the momentum equation \citep{dedner2002hyperbolic,tricco2012constrained}. This matters precisely in the regime where the magnetic field is dynamically important: a spurious $\nabla\cdot\mathbf{B}$ produces unphysical forces along the field, distorting the hydrodynamics. Obtaining both the favorable properties of Eulerian solvers and Lagrangian histories thus requires accurate Lagrangian tracer particles.

The demand for such tracers is widespread. In galaxy formation, tracers are used to separate cold- and hot-mode gas accretion \citep{kerevs2005galaxies,nelson2013moving}, to follow the cosmic-web filaments that feed halos \citep{pichon2011rigging,tillson2015angular}, as well as to track mass through the stellar and black-hole feedback cycle \citep{cadiou2019accurate}. In ISM and intracluster-medium turbulence, they are used to measure mixing times, to characterize Lagrangian statistics, and to seed dispersion in models of chemical evolution \citep{federrath2008turbulent,silvia2010dust,vazza2011turbulent}, and to advect the spectra of non-thermal particle populations such as cosmic-ray electrons \citep{winner2019evolution}.
The simplest such tracer is a massless point particle advected using the cell-centered gas velocity, typically interpolated via a cloud-in-cell (CIC) kernel. These ``classical'' or ``velocity-field'' tracers have a long history in finite-volume codes, including in \textsc{RAMSES} \citep[e.g.][]{pichon2011rigging,dubois2013blowing} and in \textsc{FLASH} \citep{dubey2012imposing}, and remain in active use \citep{nozaki2025tracking,jensen20263d}.

Classical tracers have a well-documented failure mode. As shown in \citet{genel2013following} (\textsc{AREPO}) and \citet{price2010comparison} (\textsc{FLASH}), they systematically over-concentrate in converging flows and under-sample diverging ones, with density biases that can reach an order of magnitude in turbulent and cosmological regimes. The reason is structural. In a finite-volume code, mass is redistributed between cells through the Godunov scheme, while the classical tracer is advanced only by the interpolated cell-centered velocity. As we show in Sec.~\ref{sec:theory}, these two operations are not mathematically equivalent: the latter lacks the diffusive and higher-order dispersive terms that Eulerian schemes generate. Wherever they differ significantly --- in filaments, turbulent compressions, and halo inflows --- the tracer distribution departs from the gas distribution \citep{cadiou2019accurate}. A classical tracer is an indivisible point, and so it cannot represent the diffusion that is inevitably present in the underlying fluid solver. SPH and MFM particles escape this failure only because each particle is itself the mass element, rather than a passive rider on a separate flow.
To address this failure in \textsc{AREPO}, \citet{genel2013following} replaced the deterministic velocity advection with a probabilistic jump; \citet{cadiou2019accurate} subsequently ported the scheme to \textsc{RAMSES}. In this Monte-Carlo (MC) tracer scheme, each tracer is attached to a host cell rather than to a position, and jumps to a neighboring cell at each time-step with probability $P_{i\to j} = \max(\Delta M_{ij},0)/M_i$, proportional to the Godunov mass flux. The expected tracer flux is then equal to the gas mass flux by construction, so that the Eulerian distribution of MC tracers converges to that of the gas in the $N_{\rm tracer}\to\infty$ limit. Residual per-cell noise is Poisson, with parameter $\lambda \equiv M_{\rm cell}/m_{\rm t}$ \citep{cadiou2019accurate}. MC tracers are, in this sense, correct by construction, and they are a substantial improvement over the classical scheme. 

MC tracers are not, however, without limitations. Their trajectories are fundamentally discontinuous, which excludes a differential-equation-level treatment of the tracer itself and rules out much of the standard stochastic-integration toolkit. Even when the Eulerian distribution is reproduced exactly, MC trajectories spread faster than the gas parcels they are supposed to follow \citep{genel2013following}. The data-dependent cell-exchange structure is also awkward to port to GPU architectures. Finally, there is no natural bridge from the MC jump to the continuous-trajectory Stochastic Differential Equations (SDEs) that describe other particle populations of astrophysical interest: dust and charged particles that experience drag \citep{Yan+Lazarian+Draine_2004, hirvijoki2014ascot,Moseley2025}, cosmic rays undergoing pitch-angle scattering \cite{Krumholz2022,Merten2025}, or the motion of fluid parcels subject to subgrid turbulent diffusion \citep{Marchioli2017}. These limitations motivate the present work.
An SDE is essentially an ordinary differential equation to which a random increment has been added. It thus in general has a deterministic and a stochastic part. Such processes are generally continuous, but non-differentiable. A canonical example is a drift-diffusion equation, ${\rm d}X_t = \mu\,{\rm d}t + \sigma\,{\rm d}W_t$, for which ${\rm d}W_t$ represents an infinitesimal increment of a continuous Wiener process; we revisit this formally in Section~\ref{sec:theory}. Such equations can often be rigorously mapped to (and from) continuum analogs, such as the Fokker--Planck equation in the simple case above. These equations, in analogy to ODEs, admit numerical solutions through a variety of different integrators to different orders of accuracy.\footnote{The notion of the order of accuracy is more complex in the case of SDEs than ODEs, as there are two notions. One is that of \textit{weak order}, or how accurate the scheme is \textit{statistically}, while another is that of \textit{strong order}, or how faithfully the scheme reproduces realistic trajectories. These two need not agree.}

The integration of particle SDEs is well developed in plasma physics. The \textsc{ASCOT} family of codes \citep{boozer1981monte,hirvijoki2014ascot,varje2019ascot5}, for example, follows minority species in tokamak plasmas by integrating the relevant SDEs with Euler--Maruyama or Milstein schemes, with drift and diffusion set by Coulomb collisions. This example is particularly motivating for the present study: if an SDE-based particle method is sufficient to capture particle transport in fusion devices, it should also be sufficient to capture the transport of mass in astrophysical hydrodynamics.


Outside of astrophysics, the idea of representing a fluid observable by an ensemble of Lagrangian particles whose moments match a target process has a long history. \citet{lundgren1967distribution} derived transport equations for the one-point velocity PDF in turbulent flow, and \citet{dopazo1974approach} extended the approach to the joint PDF of composition variables. \citet{pope1985pdf} unified the framework: for a wide class of turbulent reactive flows, one may construct notional Lagrangian particles evolved by a Langevin-type It\^o SDE with Gaussian noise, whose one-point PDF evolves identically to that of the real fluid parcels, provided that the drift and diffusion are moment-matched to the target. This Lagrangian-SDE approach has since become the standard tool in turbulent combustion modeling (see \citealt{haworth2010progress} for a review). To our knowledge, it has not yet been brought to bear on the problem of passive Lagrangian tracers in an astrophysical fluid code.
The method we develop in this paper, which we call the It\^o tracer, sits within this Lagrangian-SDE tradition but targets a different process in a different setting. While the ASCOT family of codes \citep[e.g.][]{hirvijoki2014ascot} describes their particle SDE methods as ``Monte-Carlo'', this name is unavailable to us, as it would cause confusion with the methods of \citet{genel2013following} and \citet{cadiou2019accurate}. We choose the name \textit{It\^o} because we want to emphasize that this is a general framework for particles whose trajectories are continuous in time, and to contrast it with the existing (discontinuous) Monte-Carlo tracer method. \textit{It\^o} highlights the continuous, SDE nature of this family of particles. However, as we will see in Sec.~\ref{sec:iton_continuity}, it is important to note that not all members of this family have sample paths that are continuous in space.


In contrast with the methods of \citet{pope1985pdf} and others who match moments to a closure-modeled turbulent flow, the It\^o tracer matches moments of the Markov transition kernel implied by the numerical flux, of which the MC tracer jump is one realization. We denote as the It\^o-$n$ tracer the continuous-trajectory tracer whose first $n$ displacement moments match those of this kernel. Matching the first two moments yields the It\^o-2 tracer; its drift is set by the Riemann-solver upwind/downwind mass flux, rather than by the cell-centered gas velocity, and its diffusion coefficient is set by the MC-jump variance. In the incompressible limit, this reduces to the first-order upwind numerical diffusion of the underlying Godunov scheme. Because the MC jump is bounded and skewed rather than Gaussian, we additionally match the third moment with a piecewise skew uniform (PSU) kick distribution, yielding the It\^o-3 tracer; this brings the method into closer agreement with the MC reference. As a by-product, the SDE framing admits a second, physically-motivated diffusion source: subgrid-scale turbulence \citep{smagorinsky1963general,lilly1966representation,wadsley2008treatment,colbrook2017scaling}. The resulting SGS-It\^o variant uses a Smagorinsky--Lilly turbulent diffusivity in place of numerical diffusion, and reduces naturally to a classical tracer in laminar flow.

The remainder of this paper is organized as follows. In Section~\ref{sec:theory}, we develop the It\^o method from the Kramers--Moyal expansion of the Markov transition kernel implied by the numerical flux, first in 1D (Section~\ref{sec:ito1d}) and then in 3D, and we introduce the SGS-It\^o variant (Section~\ref{sec:sgs_imc}). In Section~\ref{sec:square_pulse}, we validate Lagrangian fidelity using a 1D square-pulse advection test with two slope limiters (MonCen and Godunov). In Section~\ref{sec:turb}, we present a 3D decaying isothermal turbulence test at $\sigma_{\rm rms} = 15\,c_{\rm s}$, with the gas evolved in both \textsc{RAMSES} and \textsc{mini-RAMSES} to identical output, so that It\^o tracers in \textsc{mini-RAMSES} may be compared directly against reference MC tracers in \textsc{RAMSES}. We summarize our main findings and outline avenues for future work in Section~\ref{sec:summary}. In Appendix~\ref{app:finite_covariance}, we derive the complete finite-step covariance, and outline its minimal impact on relevant statistics.

\section{Theory}\label{sec:theory}
In this section, we will explain the connection between the mass continuity equation and the Markov process defined by the numerical flux, the relationship of this Markov process to the standard Monte-Carlo (MC) tracer method, and how the same Markov process may be used as a basis from which to derive a new, quasi-deterministic tracer method that we call the Itô-Monte-Carlo (It\^o) method. We first present this method in 1D for simplicity, then generalize it to 3D. After that, we present a related, alternative method, with the source of particle diffusion as sub-grid-scale (SGS) turbulence, which we term the SGS-It\^o method.

\subsection{The It\^o tracer method in 1D}\label{sec:ito1d}
In order to derive the 1D It\^o method, we must first explain how numerical diffusion and dispersion impact the 1D mass continuity equation. Following this, we will see how the Markov process implicit in the numerical flux may be used as a basis for constructing a continuous-time process with the same statistical moments. Having constructed this process, we will see that in the incompressible limit, the diffusion implied by the It\^o tracer is simply first order upwind diffusion. We finish the section by connecting numerical dispersion to the skewness. 
\subsubsection{The numerical mass continuity equation}
The 1D continuity equation for gas density may be written as:
\begin{align}
    \pder{\rho}{t} + \pder{F}{x} &= 0,\label{eq:continuity}\\
    F&\equiv \rho u.
\end{align}
Here, $\rho$ is the gas mass density, $F$ is the mass flux, $u$ is the gas velocity, $x$ is a spatial coordinate, and $t$ is time.
For a finite-volume hydrodynamical solver such as \textsc{RAMSES}, the mass-continuity equation is exactly satisfied: mass is conserved to machine precision. Inaccuracy in the solution must thus manifest through diffusive and dispersive terms, such that we may write our effective \textit{numerical} flux $F_{\rm num}$ as a series of terms:
\begin{align}
F_{\mathrm{num}} = \sum_{n=1}^\infty \frac{\partial^{n-1}}{\partial x^{n-1}}\left(\kappa_n\rho\right).
\end{align}

Equivalently,
\begin{align}
    \pder{}{x}F_{\mathrm{num}} = \sum_{n=1}^\infty \frac{\partial^{n}}{\partial x^{n}}\left(\kappa_n\rho\right).\label{eq:flux_expansion}
\end{align}
The terms $\kappa_n$ have straightforward interpretations when $n$ is sufficiently small: $\kappa_1$ corresponds to the effective advection velocity so that $\kappa_1=u$. The second coefficient $\kappa_2$ corresponds to numerical diffusion (Fokker-Planck-like). Usually, we have $\kappa_2 \simeq - u h$ for a first-order scheme, $h$ being here the grid spacing. Finally, $\kappa_3$ corresponds to a dispersive or phase-level error, and $\kappa_4$ to hyper-diffusion. Generally speaking, higher order terms are smaller and less important. A second-order scheme, like the MUSCL-Hancock scheme used in \textsc{RAMSES}, has $\kappa_2=0$ and $\kappa_3 \simeq u h^2$. This decomposition and the corresponding coefficients can be obtained via a {\it modified equation analysis} of the adopted numerical scheme.


\subsubsection{Finite-volume hydrodynamics as a Markov process}\label{sec:hydro_as_markov}
In each time-step of a finite-volume hydrodynamical solver, fluxes are passed from one cell to its neighboring cells. For stability, the CFL condition guarantees that the fraction of mass that flows from cell $i$ into neighboring cell $j$ is less than $1$. If we consider the fluid ``parcels'' within that cell to be homogenized within the cell at each timestep, with the information that we have available to us, we have no way of knowing precisely which cell the parcel goes into (nor where it came from). We can thus only know the trajectory of fluid parcels \textit{statistically}. 
Rigorously, the probability that a parcel moves from cell $i$ into cell $j$ is

\begin{align}
    P_{i\to j}\equiv\frac{1}{V_i}\int_{V_i}\mathrm{d}^3 x\int_{V_j}\mathrm{d}^3 x' q(x', t + \Delta t\mid x, t) = \frac{{\rm max}(\Delta M_{ij},0)}{M_i} .\label{eq:transition_rate}
\end{align}

where $q(x',t'|x,t)\,{\rm d}x'^3$ is the probability for a parcel to go from $x\rightarrow x'$ in a time interval ${\rm d}t$, \textit{given} it was previously at $x$ at time $t$, $V_i$ is the starting cell $i$ volume, $V_j$ the target cell volume, and $P_{i\rightarrow j}$ is the total probability that such a transition occurs. $\Delta M_{ij}$ is the mass that is transferred in a single time-step from cell $i$ to cell $j$. If it is negative, it represents an inflow of mass from $j$ to $i$. In this case, the probability is zero. This set of probabilities has no dependence on past history (the information at time $t$ is sufficient to predict the state at the next step), and so these transition probabilities form the basis of a Markov process. Note that such a Markov process only need satisfy Eq.~\ref{eq:transition_rate} in the integral sense.

In principle, there are infinitely many ways to satisfy Eq.~\ref{eq:transition_rate}. One is a discrete cell-to-cell jump: this is the basis of the MC tracer method of \citet{genel2013following} (and later, \citet{cadiou2019accurate}). Another is a continuous random walk, with drift, diffusion, and higher-moment structure chosen so that the tracer density evolves alongside the gas. This second realization is the It\^o tracer, which we derive in the remainder of this section directly from Eq.~\ref{eq:transition_rate}. By construction, both realizations generate particle fluxes that are on average proportional to the gas density fluxes, so that the accuracy of the tracers is limited only by the particle number and the accuracy of the underlying numerical scheme.

Either realization tells us that the mass continuity equation (again in 1D, for simplicity) can be thought of directly as a \textit{probability} continuity equation,
\begin{align}
    \pder{p}{t} + \pder{F_{\rm M}}{x} &= 0,
\end{align}
where $p \propto \rho_{\rm gas}$ is the probability density of a Markov process satisfying Eq.~\ref{eq:transition_rate}, and $F_{\rm M} \propto F_{\rm num}$ is the corresponding probability flux.

\subsubsection{Kramers-Moyal expansion of a Markov process}
The PDF $p(x,t)$ of a general 1D Markov process may be written using a Kramers-Moyal expansion,
\begin{align}
    \pder{}{t}p(x,t) &= \sum_{n = 1}^\infty \frac{(-1)^n}{n!}\frac{\partial^{n}}{\partial x^n}\left(B_{n}(x)p(x,t)\right).
\end{align}
The coefficients $B_n$ are given by the $n$th moment of the displacement $x'-x$ over an infinitesimal time $\epsilon$, conditional on $X_t = x$, divided by $\epsilon$ \citep[Eq.~6.99b,][]{thorne2017modern}:
\begin{align}
    B_n(x) \equiv \lim_{\epsilon \to 0}\frac{1}{\epsilon}\int (x'-x)^n\, q(x', t+\epsilon\mid x, t)\,\mathrm{d}x'.\label{eq:B_n_definition}
\end{align}

The Kramers-Moyal expansion and the modified-equation flux expansion Eq.~\ref{eq:flux_expansion} have the same structure: both describe the evolution of a density as a sum of spatial derivatives. Setting $p\propto\rho_{\rm gas}$ as required by Eq.~\ref{eq:transition_rate} and matching term by term, we immediately read off the Kramers-Moyal coefficients of the tracer process in terms of the numerical-scheme coefficients of the gas:
\begin{align}
    B_n = (-1)^{n+1}\,n!\,\kappa_n.\label{eq:B_from_kappa}
\end{align}
Eq.~\ref{eq:B_from_kappa} is the first-principles derivation of the It\^o-$n$ tracer from Eq.~\ref{eq:transition_rate}: the coefficients $B_n$ are fixed entirely by the hydro solver through its modified-equation coefficients $\kappa_n$, without reference to any particular realization of the Markov process. The accuracy of the resulting tracer method corresponds to the number of moments matched.\footnote{For the remainder of this work, we shall refer to the advection velocity $u \equiv \kappa_1$, the diffusion coefficient $\kappa \equiv \kappa_2 = B_2/2$, and the skewness, $\gamma \equiv B_3/B_2^{3/2}$ rather than $\kappa_n$ or $B_n$.}

An It\^o process is a \textit{continuous-time} random walk with a deterministic drift and a stochastic kick. We denote as the \textit{It\^o-$n$ tracer} the continuous-trajectory Markov process whose first $n$ Kramers-Moyal coefficients are those of Eq.~\ref{eq:B_from_kappa}. Its PDF is then guaranteed to evolve identically to the gas up to order $n$ in the flux expansion (Eq.~\ref{eq:flux_expansion}). \footnote{Only the $n=2$ member has almost-surely continuous sample paths, as discussed in Sec.~\ref{sec:iton_continuity}.}

In practice, evaluating the $B_n$ directly from the modified-equation coefficients $\kappa_n$ is cumbersome: the $\kappa_n$ of a given numerical scheme are typically not tabulated, and higher orders require increasingly involved modified-equation analyses. A simpler route is to use the discrete MC tracer as a computational proxy. Because the MC jump is another realization of Eq.~\ref{eq:transition_rate}, its cell-integrated moments satisfy Eq.~\ref{eq:B_from_kappa} by construction, and its single-step displacement distribution is known explicitly in terms of the Godunov mass fluxes. We can therefore read the $B_n$ off from the moments of the MC jump. In the following two sub-subsections, we follow this route to construct the It\^o-2 and It\^o-3 tracers explicitly, and then return to the general hierarchy in Sec.~\ref{sec:iton}.

\subsubsection{Constructing the Itô-2 tracer process}\label{sec:ito2}

Let the single-step displacement of the MC tracer be $\Delta X \in \{-h,0,+h\}$ with probabilities
\begin{align}
    \mathbb{P}(\Delta X = +h)&=\pR,\\
    \mathbb{P}(\Delta X = -h)&=\pL,\\
    \mathbb{P}(\Delta X = 0)&=1-\pR-\pL,\\
    \pR + \pL&\le 1.
\end{align}
$h$ is the grid-spacing, $\pR$ is the probability of jumping to the right, and $\pL$ is the probability of jumping to the left. Then the expected displacement, variance, and skewness over a time $\Delta t$ are given by
\begin{align}
    \mu\Delta t &\equiv \mathbb{E}[\Delta X], \\
    &= (\pR-\pL)\,h,\\
    \sigma^2\Delta t &\equiv \mathbb{E}[(\Delta X-\mu\Delta t)^2],\\
    &= \Big[(\pR+\pL)-(\pR-\pL)^2\Big]h^2,\\
    \gamma&\equiv \frac{\mathbb{E}[(\Delta X -\mu\Delta t)^3]}{\sigma^3\Delta t^{3/2}},\\
    &= (\pR-\pL)\frac{1 -3(\pR+\pL) + 2(\pR-\pL)^2}{\left((\pR + \pL)-(\pR -\pL)^2\right)^{3/2}}.
\end{align}
Here, $\mu$ is the mean of the process (corresponding to drift), $\sigma^2$ is the variance (corresponding to diffusion), and $\gamma$ is the skewness (corresponding to dispersion).

We may write an It\^o process that generates an overall probability density approximately proportional to the gas density. To do this, we separate the Markov process into deterministic and non-deterministic parts. Drift ($\mu$) may be treated deterministically, whereas diffusion ($\sigma^2$) must be treated stochastically. A corresponding It\^o process with the same drift and variance is,
\begin{align}
    {\rm d}X_t &= \mu\,{\rm d}t + \sigma\,{\rm d}W_t.\label{eq:ito}
\end{align}
The drift $\mu\,{\rm d}t$ is deterministic, while
${\rm d}W_t$ is an infinitesimal \textit{L\'evy increment} corresponding to a L\'evy process $W_t$. A L\'evy process such as $W_t$ has the following properties \citep{ken1999levy}:
\begin{enumerate}
    \item $W_0 = 0$ almost surely.
    \item The increments of $W_t$ are independent, e.g. $W_{t+\delta t}-W_t$ is independent of all past values $W_s$ with $s<t$.
    \item For any $s<t$, $W_t-W_s$ is equal in distribution to $W_{t-s}$.
    \item $W_t$ is stochastically continuous, i.e. for any $\varepsilon>0$, the probability that $|W_{t+\delta t}-W_t|>\varepsilon \to 0$ as $\delta t \to 0$. Sample paths are c\`adl\`ag (right-continuous with left limits).
\end{enumerate}
A Wiener process is a specific type of L\'evy process whose increments are Gaussian: $W_{t+\delta t}-W_t$ is normally distributed with mean 0 and variance $\delta t$. 
A L\'evy process has almost-surely continuous paths only when it consists of Brownian motion plus deterministic drift; any non-Gaussian L\'evy component introduces jumps. While a Wiener process is formally defined to have Gaussian increments, in practice we may opt to approximate it with any distribution whose mean is zero and variance is $\delta t$. More generally, we may approximate a L\'evy process by constructing another process that matches its statistical moments.


A process that satisfies point (iii) above is called \textit{stationary}. A careful reader may note that stationarity is not satisfied if we prescribe higher order moments ($n > 2$) to $W_t$. In this case, we must clarify that the L\'evy process is \textit{piecewise stationary}, that is, it is stationary within each hydro timestep $\Delta t$. This stationarity is critical, as it directly gives rise to the L\'evy additive property discussed in Sec.~\ref{sec:iton}, which in principle allows us to apply more general machinery in the solution of the It\^o tracer SDEs in future work.

For a Wiener process, the Fokker-Planck equation equivalent to Eq.~\ref{eq:ito} is \citep{thorne2017modern}:
\begin{align}
    \pder{p}{t} + \pder{}{x}\left(\mu p\right) &= \pdder{}{x}\left(\frac{\sigma^2}{2}p\right).
\end{align}
Thus, the advection and diffusion coefficients associated with the Markov process are, respectively,
\begin{align}
    &u = \mu = \frac{h}{\Delta t} (\pR-\pL),\\
    &\kappa = \frac{\sigma^2}{2}= \frac{h^2}{2\Delta t}\left[(\pR + \pL) -(\pR-\pL)^2\right].
\end{align}
We now note the forms that $\pR$ and $\pL$ take:
\begin{align}
    \pR &\equiv \frac{\Delta t}{\rho_{\rm c}h}{\rm max}(F_{\rm R},0),\\
    \pL &\equiv \frac{\Delta t}{\rho_{\rm c}h}{\rm max}(-F_{\rm L},0).
\end{align}
$\rho_{\rm c}$ is the cell-centered density. 

When we are in the fully compressible limit so that it may be the case that the right and left fluxes may be either both into or both out from the cell, it is useful to define
\begin{align}
    C_+&\equiv \pR + \pL &= \frac{\Delta t}{h \rho_{\rm c}}\left({\rm max}(F_{\rm R},0) + {\rm max}(-F_{\rm L},0)\right),\label{eq:cplus}\\
    C_- &= \pR -\pL&= \frac{\Delta t}{h \rho_{\rm c}}\left({\rm max}(F_{\rm R},0) - {\rm max}(-F_{\rm L},0)\right).\label{eq:cminus}
\end{align}
These are two (possibly different) CFL-like numbers associated with diffusion and advection, respectively.\footnote{Note that $C_- \in [-1,+1]$ and $C_+ \in [0,1]$.} Thus:
\begin{align}
    u &= \frac{h}{\Delta t}C_-,\label{eq:drift}\\
    \kappa &= \frac{h^2}{2 \Delta t}(C_+-C_-^2),\label{eq:diffusion}\\
    \gamma& = C_-\frac{1-3 C_+ + 2C_-^2}{(C_+ - C_-^2)^{3/2}}\label{eq:skewness}
\end{align}
We have included the skewness in terms of $C_\pm$ for completeness and reference in subsequent sections.
\subsubsection{The incompressible limit: first order upwind diffusion}
In the incompressible limit, $|C_-| = C_+ \equiv C$, in which case $u$ becomes:
\begin{align}
    u = \frac{F_{\rm D}}{\rho_{\rm c}},
\end{align}
where $F_{\rm D}$ is the flux \textit{downwind} of the cell-center, and $\kappa$ becomes
\begin{align}
    \kappa &= \frac{h}{2}|u| - \frac{\Delta t}{2}|u|^2,\\
    &= \frac{1}{2}h|u|\left(1-\frac{\Delta t |u|}{h}\right),\\
    &= \frac{h^2}{2\Delta t}C\left(1-C\right),\label{eq:fo_upwind_diff}
\end{align}
$C$ is the Courant-Friedrichs-Lewy number defined by this newly defined downwind advection velocity $u$. Note that this diffusion is exactly the numerical diffusion of a first order upwind scheme, provided that the fluxes are also from a first order upwind scheme. 
\subsubsection{Constructing the Itô-3 tracer process}\label{sec:ito3}
In the previous section, we constructed the It\^o-2 tracer, which matches the first two moments of the transition kernel. It sets the deterministic drift $B_1=u$ and diffusion coefficient $B_2/2=\kappa$, correctly reproducing the advective and diffusive terms of the underlying numerical scheme, but not its dispersion. The It\^o-3 tracer additionally matches the third Kramers--Moyal rate coefficient $B_3$, or equivalently the skewness in Eq.~\ref{eq:skewness}, where
\begin{align}
    B_3 \equiv \frac{\mathbb{E}\left[(\Delta X-\mu\Delta t)^3\right]}{\Delta t} = \frac{h^3}{\Delta t}\,C_-\left(1 - 3C_+ + 2C_-^2\right).\label{eq:B3}
\end{align}

Like the first and second cumulants, $B_3$ is a function of the properties of the numerical solution provided by the hydro solver. For a tracer step $\delta t \le \Delta t$, the process has skewness,
\begin{align}
    \gamma(\delta t) &= \frac{\mathbb{E}[(\Delta X_{\delta t} - u\,\delta t)^3]}{(2\kappa\,\delta t)^{3/2}} = \frac{B_3}{(2\kappa)^{3/2}}\delta t^{-1/2} \\
    &= \gamma(\Delta t)\left(\frac{\delta t}{\Delta t}\right)^{-1/2}\label{eq:sub_skewness}
\end{align}

Again, $u$ is given by Eq.~\ref{eq:drift}, $\kappa$ is given by Eq.~\ref{eq:diffusion}, and $\gamma(\Delta t)$ is given by Eq.~\ref{eq:skewness}. 

\begin{figure}
    \centering
    \includegraphics[width=\columnwidth]{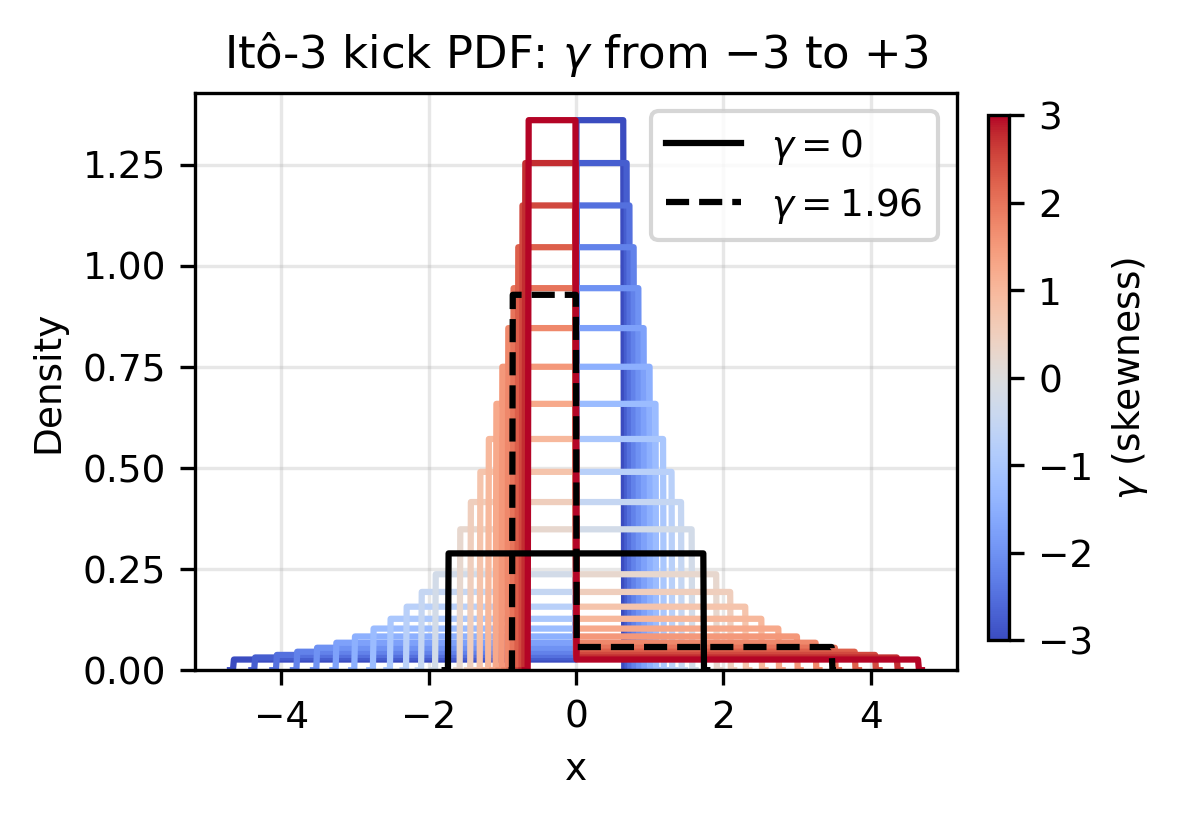}
    \caption{The Piecewise Skew Uniform (PSU) distribution for varying skewness $\gamma$. The distribution has mean zero and variance one regardless of the level of skewness. The distribution can be seen to be antisymmetric with respect to the sign of the skewness. The uniform distribution (solid black) is shown for comparison. In the limit of $\gamma=0$ the PSU is equivalent to it. We also show the distribution corresponding to the typical skewness in our square pulse advection test (black, dashed; cf. Sec.~\ref{sec:square_pulse}, Figs.~\ref{fig:square_pulse0}-\ref{fig:trajectories}).}
    \label{fig:psu}
\end{figure}

We construct a distribution that matches this next order moment, which we refer to as the piecewise skew uniform (PSU) distribution. The PSU has a mean of zero, variance of 1, and prescribed skewness $\gamma$. The distribution extends $-a_-$ to the left of zero, $+a_+$ to the right, with heights $b_-$ and $b_+$ to the left and right of zero, respectively. For a given skewness $\gamma$, these are,
\begin{align}
    a_\pm &= \frac{1}{3}\left(\mp2\gamma + \sqrt{27 + 4 \gamma^2}\right),\\
    b_\pm &= \frac{1}{9}\left(\pm 2\gamma + \frac{1}{2}\left(\frac{27 + 8\gamma^2}{27+4\gamma^2}\right)\sqrt{27+4\gamma^2}\right).
\end{align}
A plot of this distribution for varying skewness is shown in Fig.~\ref{fig:psu}. Thus, the It\^o-3 tracer matches the behavior of the gas up to the third order dispersive term, and evolves according to the following process:
\begin{align}
    {\rm d}X_t = u\,{\rm d}t + \sqrt{2\kappa}\,{\rm d}W_t,
\end{align}
where $u$ and $\kappa$ are defined according to Eqs.~\ref{eq:drift} and \ref{eq:diffusion}, and the stochastic increment $W_t$ is a L\'evy process with its increments PSU distributed with mean $0$, variance $\delta t$, and skewness $\gamma(\delta t)$.

\subsubsection{The It\^o-$n$ tracer hierarchy}\label{sec:iton}

As briefly mentioned in the previous section, the It\^o-2 and It\^o-3 tracers are the first two members of a hierarchy. For general $n$, Eq.~\ref{eq:B_from_kappa} defines the It\^o-$n$ tracer as the Markov process whose first $n$ Kramers-Moyal coefficients are fixed by the modified-equation expansion of the hydro scheme (Sec.~\ref{sec:hydro_as_markov}). For $n=2$ and $n=3$, we evaluated $B_1=u$, $B_2=2\kappa$, and $B_3$ (Eq.~\ref{eq:B3}) from the mean, variance, and third central moment of the MC single-step displacement, divided by the hydro timestep $\Delta t$.

A key property of this construction is that the coefficients $B_k$ are defined as rates \textit{per unit time}, carrying no explicit timestep of their own. They thus describe an SDE for a continuous process across any subinterval $\delta t \le \Delta t$, with an increment $\Delta X_{\delta t}$ whose mean and next $n-1$ central moments scale linearly with $\delta t$:

\begin{align}
    \mathbb{E}[\Delta X_{\delta t}] &= B_1\,\delta t =u\,\delta t,\\
    \mathbb{E}\!\left[(\Delta X_{\delta t} - u\delta t)^2\right] &= B_2\,\delta t = 2\kappa\,\delta t,\\
    \mathbb{E}\!\left[(\Delta X_{\delta t} - u\delta t)^3\right] &= B_3\,\delta t,\\
    &\,\,\vdots \nonumber\\
    \mathbb{E}\!\left[(\Delta X_{\delta t} - u\delta t)^n\right] &= B_n\,\delta t.
\end{align}
This is the L\'evy additive property: aggregating $N$ substeps of size $\delta t$ yields the same first $n$ central moments as a single increment of size $N\delta t$. The tracer substep is, to order $n$ (and so long as $\delta t \le \Delta t$), free. This means that taking $N$ substeps of size $\delta t = \Delta t/N$ is equivalent to taking a single step of size $\Delta t$. This is guaranteed by the above linearity. This does not necessarily imply that there is no added benefit in taking smaller substeps: the higher order terms may have error that decreases super-linearly.

We used the L\'evy additive property implicitly in writing the It\^o-3 skewness at finite substep (Eq.~\ref{eq:sub_skewness}). Extending the hierarchy to $n \ge 4$ would require matching hyper-diffusion and higher numerical-flux terms, which are subdominant for the second-order MUSCL-Hancock scheme used in \textsc{RAMSES}.

\subsubsection{Path continuity beyond second order}\label{sec:iton_continuity}
Beyond second order, a continuous-time process does not generically have continuous sample paths. The It\^o-2 tracer is an ordinary diffusion driven by a Wiener process: its paths are almost surely continuous, and its Kramers--Moyal coefficients $B_k$ vanish for $k\ge3$. Pawula's theorem implies a Markov process whose Kramers--Moyal expansion has non-zero coefficients above second order cannot terminate at finite order while preserving non-negative probability \citep{pawula1967approximation,risken1989fokker}.

This means that a nonzero $B_{k\ge 3}$ can therefore not be accommodated by \textit{only} introducing a skewness or other higher moment to a continuous particle trajectory. In the Markov framework, it necessitates a jump component, and thus according to Pawula, an infinite hierarchy of Kramers--Moyal coefficients. This means that for a process with $B_k\neq 0$ for $k\ge 3$, the paths are c\`adl\`ag: right-continuous with left limits, and continuous \textit{between} jumps. It's thus worth emphasizing that the label $n$ in It\^o-$n$ means that we match the first $n$ Kramers-Moyal terms, rather than that the order of termination of the series. In practice this is not so critical, as each term will tend to be smaller than the last. What \textit{is} important is to note that the It\^o-$n\ge3$ tracers do not represent stochastic processes with continuous sample paths. They do follow the gas, but offer less philosophical advantage over using a traditional MC tracer scheme, although they may be more accurate in a general 3D, turbulent setting.


\subsection{Generalizing It\^o tracers to 3D}\label{sec:ito3d}

In order to generalize the It\^o method to 3D, we again return to the Markov transition kernel as a baseline. 
As before, the advection term is simply the expected displacement divided by the timestep. This involves no cross terms and so may be defined per dimension exactly as in Eq.~\ref{eq:drift}.

Because the Markov process only ever moves in exactly one direction per single-step transition (a property realized directly by the MC tracer), the matrix is diagonal (i.e. if the particle moved in $x$, we know with certainty it did not move in $y$). Its raw second moment is therefore,
\begin{align}
    R_{ij}\equiv\mathbb{E}[\Delta X_i\Delta X_j]
    =\mathbb{E}[\Delta X_i^2]\delta_{ij}.\label{eq:raw2nd}
\end{align}

Because the Markov process only ever moves in exactly one direction per single-step transition (a property realized directly by the MC tracer), the matrix is diagonal (i.e. if the particle moved in $x$, we know with certainty it did not move in $y$). A subtlety here is that this differs from the central covariance at finite time-step, and it is the central covariance (rather than the raw second moment) that determines the rate coefficient $B_2$. This has proven to be inconsequential to the results and performance of the tracer, but we elaborate on this more in Appendix~\ref{app:finite_covariance} for completeness.

We retain the finite-step variance along each coordinate but neglect the cross covariances, giving the diagonal diffusion tensor
\begin{align}
    \kappa^{\rm diag}_{ij}
    &=\frac{h_i^2}{2\Delta t}
      \left(C_{+,i}-C_{-,i}^2\right)\delta_{ij}.
      \label{eq:3d_kappa_diag}
\end{align}

This tensor matches the MC mean and variance along each coordinate, rather than the complete covariance of the displacement vector. The 3D It\^o SDE is
\begin{align}
    {\rm d}{\mathbf X}_t &= \gvel \,{\rm d}t + {\bm \Sigma}\cdot{\rm d}{\mathbf W}_t,\label{eq:imc_sde}\\
    u_i &\equiv \frac{h_i}{\Delta t}C_{-,i},\label{eq:3d_drift}\\
    {\bm\Sigma}{\bm\Sigma}^{\rm T} &= 2{\bm\kappa}.\label{eq:3d_diff}
\end{align}

For Eq.~\ref{eq:3d_kappa_diag}, $\Sigma_{ij}=\sqrt{2\kappa^{\rm diag}_{ii}}\,\delta_{ij}$ and the kick components are independent. Equations~\ref{eq:cplus} and \ref{eq:cminus} define $C_{\pm,i}$ along each coordinate $i$. For the It\^o-3 tracer, each $W_{t,i}$ also matches the one-dimensional third central moment and hence the marginal skewness in Eqs.~\ref{eq:skewness} and \ref{eq:B3}; we do not match mixed third moments.

\subsection{Sub-grid scale diffusion for It\^o tracers: a physically-motivated alternative}\label{sec:sgs_imc} 
The It\^o framework is general enough that we can design another form of tracer that is more physically motivated. Implemented into \textsc{mini-RAMSES} is a subgrid-scale (SGS) turbulence model \citep{Kretschmer2020}, which evolves according to,
\begin{align}
  \frac{\partial (\rho e)}{\partial t} &+ \nabla\cdot(\rho e\,\mathbf{u})
  = \nonumber\\
  &-\frac{2}{3}\,\rho e\,\nabla\cdot\mathbf{u}
  + \frac{\SLconst^4}{2}\rho\,h\,\sigma_{\rm turb}\,\phi_{\mathrm{diss}}\, - \frac{\sigma_{\rm turb}}{h}\,\rho e
\end{align}
where the subgrid turbulent velocity is,
\begin{equation}
  \sigma_{\rm turb} = \sqrt{2\, e}
\end{equation}
 with $e$ the specific subgrid turbulent energy. The viscous heating rate $\phi_\mathrm{diss}$ is,
\begin{equation}
  \phi_{\mathrm{diss}} = 2\, S_{ij} S_{ij} - \frac{2}{3}\, (\nabla\cdot\mathbf{u})^2,
\end{equation}
the strain-rate tensor $S_{ij}$ is,
\begin{equation}
  S_{ij} = \frac{1}{2}\left( \frac{\partial u_i}{\partial x_j} + \frac{\partial u_j}{\partial x_i} \right),
\end{equation}
and $\SLconst$ is the Smagorinsky--Lilly constant \citep[usually chosen to be between 0.1 and 0.2;][]{lilly1966representation}, and $h$ is the grid cell size.

We initialize with the stationary solution in the case of decaying turbulence,
\begin{equation}
  \rho e\big|_{\mathrm{eq}} = \rho\,h^2\,\phi_{\mathrm{diss}}\,\frac{\SLconst^4}{2}
  \quad \Rightarrow \quad
  \sigma_{\rm turb} = \SLconst^2\,h\,\sqrt{\phi_\mathrm{diss}}.
\end{equation}

With this subgrid turbulence model in hand, we may define the SGS turbulent diffusion coefficient \citep{smagorinsky1963general, lilly1966representation, wadsley2008treatment, colbrook2017scaling},
\begin{align}
  \kappa_{\rm turb} &\equiv h\,\sigma_{\rm turb}
\end{align}
Because this corresponds to a Fickian diffusion rather than a Fokker-Planck style diffusion, we must add a gradient term to correct for the fact that the particles will naturally tend to preferentially move away from high diffusivity regions and toward low diffusivity regions:
\begin{equation}
  {\rm d}\mathbf{X}_t = \bigl( \mathbf{u} + \nabla\kappa_{\rm turb} \bigr)\, {\rm d}t + \sqrt{2\kappa_{\rm turb}}\, {\rm d}\mathbf{W}_t.
\end{equation}
Here, $\gvel$ is the local fluid velocity (cell-centered in our scheme, rather than the flux-defined velocity in Eq.~\ref{eq:drift}), and $\mathbf{W}_t$ is a vector of independent Wiener processes. Note that with this method, if a flow is laminar, there will tend to be little to no subgrid turbulence, and thus these SGS-It\^o particles will behave as classical tracer particles, as they use the same advection velocity.

The continuum equation that this SDE is equivalent to is:
\begin{align}
    \pder{p}{t} + \nabla\cdot(p\gvel) &= \nabla\cdot(\kappa_{\rm turb}\nabla p),
\end{align}
for a probability density of particles $p$.

\section{Implementation algorithm and tests}
\subsection{Description of algorithm}\label{sec:algorithm}
Within each cell of the simulation, we compute the quantities in Eqs.~\ref{eq:drift}-\ref{eq:skewness}. We interpolate these cell-based quantities onto particle positions using the cloud-in-cell \citep[CIC,][]{birdsall1969clouds} interpolation kernel. At a cell center, this interpolation reproduces the moments of the traditional MC algorithm along each coordinate.

The SDE (Eq.~\ref{eq:imc_sde}) is then solved with the Euler-Maruyama method \citep{kloeden1992numerical}. For a non-infinitesimal timestep $\delta t> 0$, we thus update the particle's position from time $n$ to $n+1$ via:
\begin{align}
    {\mathbf X}^{n+1} &= {\mathbf X}^n+\gvel \,\delta t + {\bm \Sigma}\cdot{\mathbf W}_{\delta t},
\end{align}
where $\gvel$ and ${\bm \Sigma}$ are given by Eqs.~\ref{eq:3d_drift} and \ref{eq:3d_diff}, respectively, and ${\bm W}_{\delta t}$ is uniformly distributed for the It\^o-2 tracer, and PSU distributed for the It\^o-3 tracer.\footnote{Here and throughout this paper, we opt for the simplest case with the particle step size $\delta t = \Delta t$, and leave sub-stepping and other variable-timestep techniques for later work.} In each case, it has per-dimension variance $\delta t$ and (for the It\^o-3 tracer) anisotropic skewness as implied in each dimension by the MC tracer process (Eqs.~\ref{eq:skewness},~\ref{eq:B3},~\ref{eq:sub_skewness}). Quantities here without superscripts are assumed to be calculated at time $n$.

A conceptual advantage of the present framework, relative to the MC tracer, is that tracer evolution is by construction a standard numerical-integration problem for an It\^o SDE. Any stochastic integrator may therefore be substituted for Euler-Maruyama in Eq.~\ref{eq:imc_sde}: higher strong-order schemes such as Milstein's method \citep{milstein1975approximate} or stochastic Runge-Kutta, implicit variants for stiff terms, or exponential integrators when the coefficients are locally smooth. This flexibility is inaccessible to the MC tracer, whose discrete jump is not itself the solution of a differential equation. For the tests presented here we have found Euler-Maruyama sufficient; the leading error is dominated by moment-matching to the MC jump rather than by the temporal discretization of the resulting SDE.

While there is no formal stability requirement on this algorithm per se, for accuracy of the solution, one may wish to adhere to a ``stay local'' condition, where particles do not tend to go far away in the next tracer step. In the case of an unbounded ${\mathbf W}_{\delta t}$, this can only be true in a statistical, rather than absolute sense. One might require that the expected r.m.s. displacement is
\begin{align}
    \langle (u_i \delta t + \Sigma_{ij}W_{\delta t,j})^2\rangle^{1/2} < h
\end{align}
in each dimension $i$. For the diagonal $\Sigma_{ij}=\sqrt{2\kappa_i}\delta_{ij}$ used in our simulations, this condition becomes
\begin{align}
    \delta t < \frac{\kappa_i}{u_i^2}\left(\sqrt{1 + \frac{h^2 u_i^2}{\kappa_i^2}}-1\right).
\end{align}
For the SGS-It\^o tracer with isotropic diffusion $\kappa_{\rm turb} = h \sigma_{\rm turb}$, 
\begin{align}
    \delta t < \frac{h}{u_i^2}\left(\sqrt{\sigma_{\rm turb}^2 + u_i^2}-\sigma_{\rm turb}\right).
\end{align}
For SGS-It\^o turbulence simulations, we adhere to this tracer-step restriction (Sec.~\ref{sec:turb}).

For the It\^o tracer with $\kappa_i$ given in Eq.~\ref{eq:diffusion}, with our choice of $\delta t = \Delta t$ (footnote above) and the $\Delta t^{-1}$ scaling of $\kappa_i$, this simplifies substantially, giving us the standard
\begin{align}
    \frac{u_i\Delta t}{h} < 1.
\end{align}
This is required for stability of the hydrodynamics regardless. 

Another, more stringent condition might be that the \textit{maximum} displacement of the tracer particle over one tracer step must be less than one cell. For the uniform distribution, this has a well defined answer. We note that $W_{\delta t} \in [-\sqrt{3 \delta t},+\sqrt{3\delta t}]$ in this case, which gives an accuracy criterion of:
\begin{align}
    \frac{u_i\delta t}{h} < \frac{1}{4}.
\end{align}

While we adhere to the latter criterion, we have found that it makes little difference to the results. Further, for the It\^o-3 tracer, a non-zero skewness will make the theoretical maximum displacement larger, and this criterion should be replaced by an even more restrictive one. Despite this, the method appears robust and shows performance comparable to the MC tracer (e.g. Figs.~\ref{fig:square_pulse0}-\ref{fig:lagrangian_fidelity}).

\subsubsection{Modification for the SGS-It\^o tracer}
The details of the SGS-It\^o tracer method described in Sec.~\ref{sec:sgs_imc} are slightly different. First, the stochastic increment ${\mathbf W}_t$ is isotropic and symmetric with skewness $\gamma = 0$. Second, as the diffusion here is Fickian rather than Fokker-Planck, one also needs to compute a gradient of $\kappa_{\rm turb}$. We do this via the standard technique used in smoothed-particle hydrodynamics where one takes the gradient of the kernel. 

This says that, for a given kernel $\phi$ and a field quantity $f$,
\begin{align}
    f(\mathbf{x}_p) &= \sum_{s} f_s\phi_s(\mathbf{x}_p),\\
\nabla f &= \sum_{s}f_s \nabla \phi_s(\mathbf{x}_p) .
\end{align}
${\mathbf x}_p$ is the position of the particle, and the sum is over all cells $s$ within the kernel's stencil. In our case, the CIC kernel is a product of three 1D linear interpolations, so the gradient is a linear interpolation of first order finite difference derivatives. 

\subsection{A guide to implementation of the Itô tracer}

In this subsection, we briefly go over a recipe for how to implement It\^{o} tracers into a general, grid-based finite-volume code. If a code already has MC tracers implemented, as well as classical tracers, then the It\^{o} tracer directly follows. The key components are as follows:

\paragraph*{Face-centered flux storage.} As can be seen in Eqs.~\ref{eq:drift}, \ref{eq:diffusion}, and \ref{eq:skewness}, the It\^o method relies upon tracer particles having access to this information. The simplest way to implement this is to store the fluxes from the Riemann solver into a separate array (or else, something equivalent, such as $\pR$, $\pL$). This is an additional, \textit{six-dimensional} array per cell. This need is not distinct from that of the MC tracer. It is worth noting that the SGS-It\^o tracer needs only one additional field (albeit a more expensive one, as it requires to solve one additional equation), the subgrid turbulence level $\sigma_{\rm turb}$. 

\paragraph*{Particle interpolation method.} While one can, in principle, use any method to interpolate the above field variables onto particle positions, we prefer the cloud-in-cell (CIC) method. At a cell center, this interpolation reproduces the moments of the traditional MC algorithm along each coordinate. This is also true of the nearest-grid-point (NGP) method, although that method is much noisier by nature, and so we do not prefer it \citep{achikanath2024ahkash, federrath2025computational}.

\paragraph*{Random number generation.} It\^o tracers require a reproducible source of noise that is uncorrelated between particles. There are many ways to do this. We use a Fortran implementation of Pierre L'Ecuyer's RngStreams library, seeded by a specified random seed in the \textsc{mini-RAMSES} namelist file. At startup each MPI rank advances its copy of the specific It\^o tracer stream so ranks draw from non-overlapping subsequences. During each kick-drift pass, every tracer consumes the next draws from that stream. Successive draws act as independent random variables for the SDE; the sequence is deterministic given seed and processing order. This is a serial stream, not per-particle or per-cell RNG addressing, although that is another viable option.

\paragraph*{Random kick distribution.} A central part of the methods described in this paper is that they represent diffusion through random kicks. There is a rigorous map between the continuum, Fokker-Planck diffusion and the Itô random kicks so long as the random kicks are \textit{gaussian}. While this is rigorously true, practically, we find it to be of little consequence, and a cheap alternative that produces near-indistinguishable results is the uniform distribution, used throughout this paper for the It\^o-2 tracer. 

For each hydro-step, the algorithm is thus roughly,
\begin{enumerate}
    \item Store hydro fluxes.
    \item Interpolate drift, diffusion, and optional higher moments onto the particle position.
    \item Generate random numbers.
    \item Apply the appropriate It\^o positional kick using those random numbers.
\end{enumerate}

The details may differ based upon the specific kick-drift and SDE integration algorithms chosen.

\subsection{Test of Lagrangian fidelity: Linear advection of a square pulse}\label{sec:square_pulse}
\begin{figure}
    \centering
    \includegraphics[width=1.0\columnwidth]{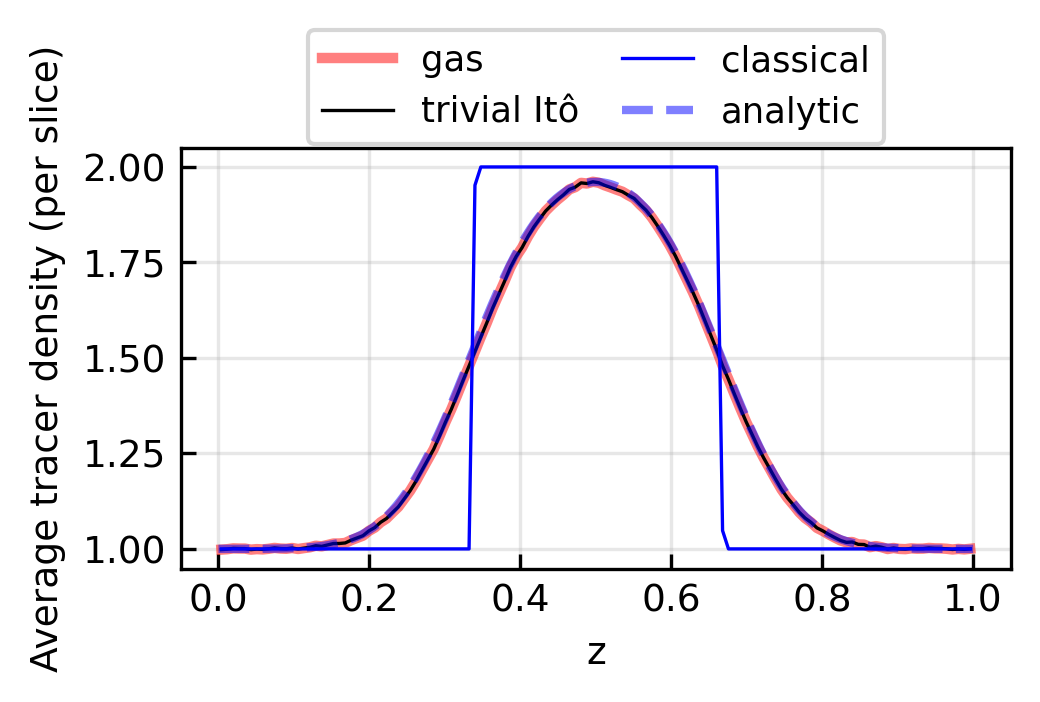}
    \caption{The square pulse advection test using the first-order Godunov method, presented against classical tracers and a ``trivial'' It\^o tracer with constant advection and diffusion (no interpolation of gas quantities). Initial conditions are as described in Sec.~\ref{sec:square_pulse}. The square pulse has been advected to the right for one period. The effective particle count here is between $262,144$ and $524,288$ particles per cell, as this is a 3D test where we average over slices to obtain the tracer statistics. The classical tracer particle mostly preserves the initial condition, while the trivial It\^o particles closely follow the gas (and the pictured analytic solution). Without random kicks to provide diffusion, the classical tracers cannot perfectly follow the gas, even in a simple 1D test such as this. 
}
    \label{fig:square_pulse0}
\end{figure}

We now present our first test of our It\^o tracer particle algorithm. The test is an advection of a square pulse in density with an isothermal equation of state (shown in Figs.~\ref{fig:square_pulse0},~\ref{fig:square_pulse}). The initial conditions for this test are:
\begin{align}
    P_0 &= 1, \label{eq:ic1}\\
    \gvel_0 &= \hat{\mathbf z},\label{eq:ic2}\\
    \rho &= 2 \qquad z \in \left(\frac{1}{3},\frac{2}{3}\right),\label{eq:ic3}\\
    \rho &= 1 \qquad {\rm otherwise}\label{eq:ic4}
\end{align}
We use an isothermal equation of state. The boundary conditions are periodic, and we run the test for one advection period. We double the number of particles within the dense region to match the ICs, and give the particles each a fixed mass. We use the Harten-Lax-van Leer contact (HLLC) approximate Riemann solver. 

We perform two different tests: a simulation using the first-order Godunov scheme \citep[piecewise constant, no reconstruction at all within each cell;][]{godunov1959finite}, and a simulation using the second-order MUSCL Hancock scheme with the Monotonized Central slope limiter \citep[MonCen;][]{van1977towards}. The fluxes from each of these  methods are then fed into the following tracer particle algorithms:
\begin{enumerate}
    \item Classical tracer
    \item MC tracer
    \item It\^o-2 tracer (uniformly distributed $W_t$)
    \item It\^o-3 tracer (PSU distributed $W_t$).
\end{enumerate}
As a reminder, the classical tracer simply interpolates cell-centered gas velocity, the MC tracer probabilistically jumps from cell to cell, and the It\^o methods have a deterministic advection with a random kick corresponding approximately to numerical diffusion. All but the classical tracer are as described in Sec.~\ref{sec:theory}. We do not test the SGS-It\^o method with this test, because with a uniform velocity, it reduces to being equivalent to the classical tracer (the SGS turbulent diffusivity will remain zero, or very nearly zero). 

Although this is a 1D test, we run it in full 3D with a resolution of $N_x = 128$ grid cells per dimension, and $16$ particles per cell in the low density region and $32$ in the high density region. That means that when averaging over slices in the $z$-direction, we have an effective particle count of between $16\times128^2 = 262,144$ and $32\times128^2 = 524,288$ per slice. This should produce a relative Poisson density error of $<0.2\%$. We choose this high particle count to highlight differences between the methods. 

We also test a ``trivial'' It\^o tracer (Fig.~\ref{fig:square_pulse0}), which has a constant velocity and diffusion coefficient, utilizing no local fluid information at all. The trivial particle velocity is $\gvel = \hat{\mathbf z}$ and the diffusion is chosen as $\kappa = C(1-C)h^2/(2\Delta t)$, the standard first order upwind diffusion coefficient (Eq.~\ref{eq:fo_upwind_diff}). The $C$ here is not equal to the actual CFL number (which includes the sound speed), but is only one due to advection: $C = u\Delta t/h.$ This is the same value as the baseline value in Fig.~\ref{fig:square_pulse_deviations} (middle panel). This illustrates that the main difference between the It\^o tracer as we have defined it in Eqs.~\ref{eq:drift}-~\ref{eq:skewness} and the trivial It\^o tracer with constant advection and diffusion are the fact that the advection velocity and diffusion vary across the density interface. Without this variation, the It\^o tracers in Fig.~\ref{fig:square_pulse} would have the same distribution as the trivial It\^o tracers in Fig.~\ref{fig:square_pulse0}, i.e. they would show the distribution equivalent to a first order scheme. 

\begin{figure}
    \centering
    \includegraphics[width=1.0\columnwidth]{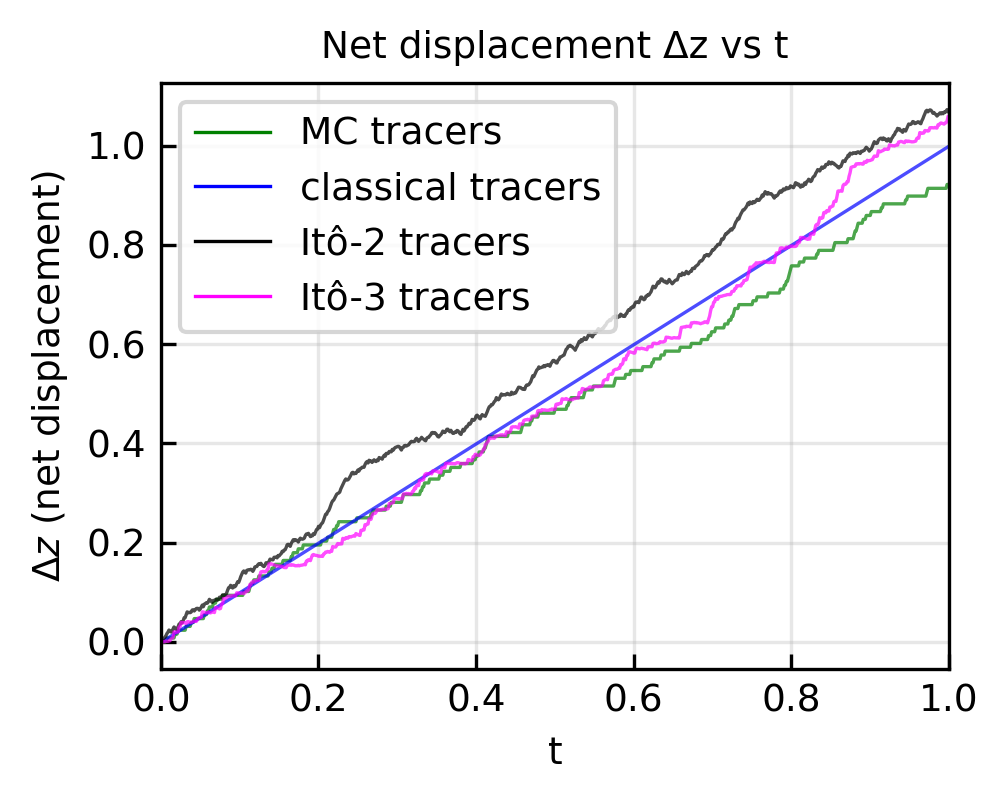}
    \caption{Particle trajectories for the MC tracer, classical tracer, and It\^o-2 and It\^o-3 tracers for the square pulse advection test described in Sec.~\ref{sec:square_pulse}. As the classical tracer (blue) simply interpolates cell-centered velocities, it uniformly proceeds from left to right. The MC tracer (green) can only stay put or jump to the right at each timestep in this test, and so its trajectory looks like a staircase with random heights and lengths. The It\^o-2 tracer exhibits more standard random-walk behavior, while the It\^o-3 tracer begins to look more like the MC tracer. This is because for this test, the PSU kick of the It\^o-3 tracer has a large probability of a small-amplitude movement to the left and a small probability of a large-amplitude movement to the right (cf. Fig.~\ref{fig:psu}).}
    \label{fig:trajectories}
\end{figure}

In order to give the reader a better picture of what these particles do in practice, we have also plotted the trajectories of a single particle using each of these tracer schemes in Fig.~\ref{fig:trajectories}. The classical tracer traces out a straight line, as the cell-centered gas velocity is constant in the simulation, while the other tracers follow their own flavors of random walk. The It\^o-2 tracer shows a more traditional random walk behavior, while the It\^o-3 tracer visually looks more similar to the MC tracer, tending to jump farther to the right than to the left. This is consistent with the distribution in Fig.~\ref{fig:psu}.

\subsubsection{Diffusion of a sheet}\label{sec:sheet}
\begin{figure*}
    \centering
    \includegraphics[width=1.0\textwidth]{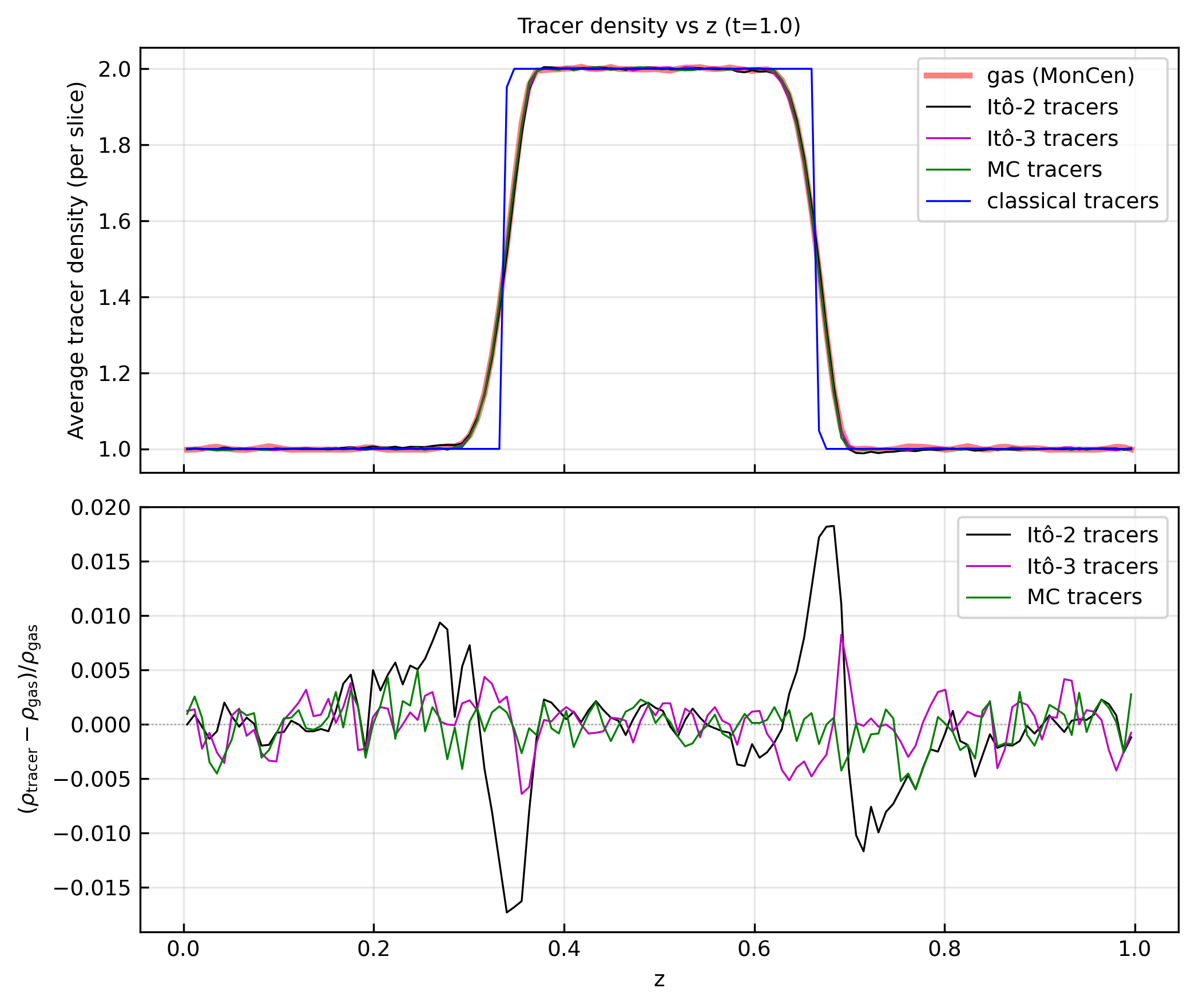}
    \caption{ {\bf Top:} The same square pulse test as Fig.~\ref{fig:square_pulse0}, but using a second-order MUSCL-Hancock scheme with the MonCen slope limiter, and depicting the classical, MC, It\^o-2, and It\^o-3 tracer methods. The classical tracer particle mostly preserves the initial condition, while the other methods (It\^o, MC) closely follow the gas. {\bf Bottom:} The same tracer curves as in the top panel, but shown after subtracting off and dividing by the gas density, to highlight errors in each of the methods. We do not show the classical tracer here, as the errors are much larger. While the It\^o-2 tracer (black) provides a much better fit than the classical tracer, it can be seen that the particles tend to first dip, then overshoot in density leading into the pulse, and do the reverse upon leaving the pulse; this is a form of numerical bias, albeit significantly reduced compared to classical tracers. The It\^o-3 tracer (magenta), which additionally matches the skewness of the MC tracer process, significantly reduces these effects, leading to better agreement with both the gas and the MC tracer.  }
    \label{fig:square_pulse}
\end{figure*}

\begin{figure}
    \centering
    \includegraphics[width=1.0\columnwidth]{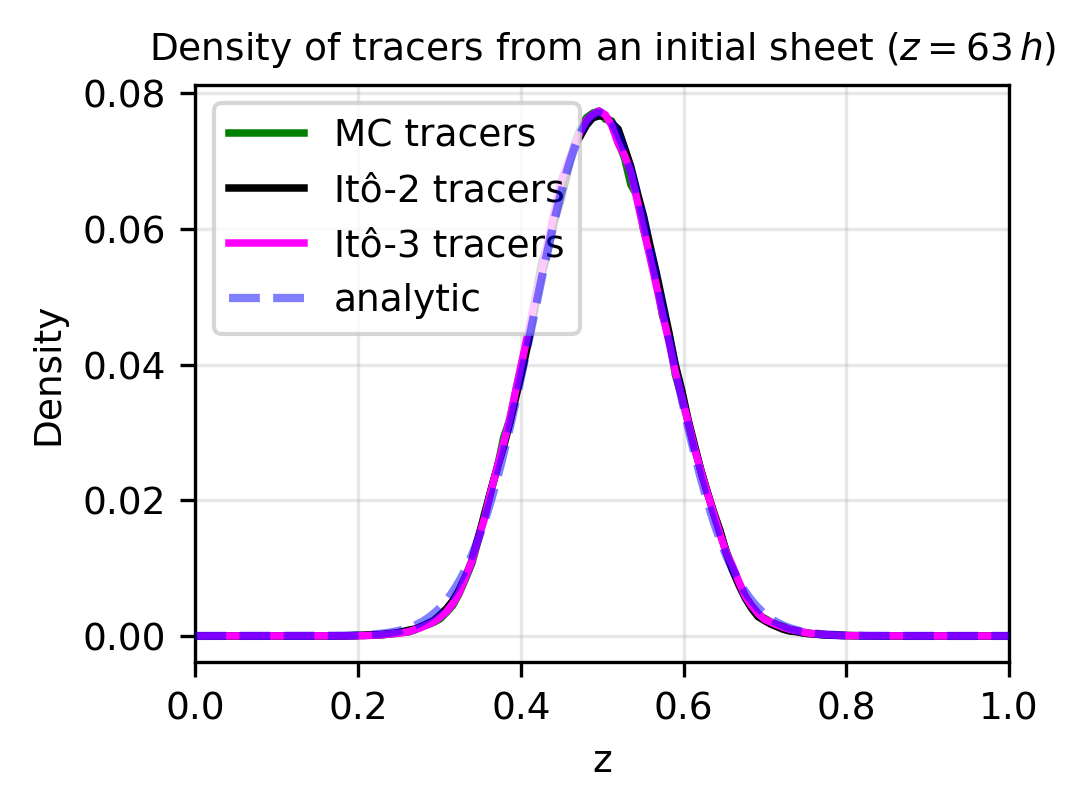}
    \caption{The distribution of particles initially located in a sheet located from $z = 63h \rightarrow 64h$ after one advection period. These particles are part of the square pulse test in Sec.~\ref{sec:square_pulse}. The distribution of all stochastic particle types is approximately equal to the analytic Gaussian, differing mainly at the tails where the advection and diffusion coefficients vary within the transition regions of the pulse from high to low density or vice versa. Classical tracers are not pictured, as they maintain the initial condition, staying within a sheet.}
    \label{fig:lagrangian_fidelity}
\end{figure}

In addition to closely comparing the global density distribution of each tracer method (Sec.~\ref{sec:stat_comparison}, Fig.~\ref{fig:square_pulse}), we examine the spread of particles initially located in a sheet between $z = 63 h$ and $64h$, thus centered at $z = 63.5h = 0.49609375$. This amounts to $N= 32\times 128^2 = 524,288$ particles. The results of this can be seen in Fig.~\ref{fig:lagrangian_fidelity}.

The Godunov first order method is equivalent to an advection-diffusion equation. We may thus write an analytic expression for the first order solution in terms of its Green function. In the frame that uniformly translates in space with the fluid:
\begin{align}
    G(z,z';t) &\equiv \frac{\exp\left(\frac{-(z-z')^2}{4 \kappa t}\right)}{\sqrt{4\pi\kappa t}},\label{eq:green_function}
\end{align}
with $\kappa$ given by Eq.~\ref{eq:fo_upwind_diff}. This is only true for a solution with open boundaries, although at sufficiently early times, this should still be accurate. This solution is what is labeled as ``analytic'' in Fig.~\ref{fig:lagrangian_fidelity}. We also convolve Eq.~\ref{eq:green_function} with the initial conditions (Eqs.~\ref{eq:ic1}-\ref{eq:ic4}) to obtain the analytic solution shown in the bottom panel of Fig.~\ref{fig:square_pulse}. 

\begin{figure}
    \centering
    \includegraphics[width=1.0\columnwidth]{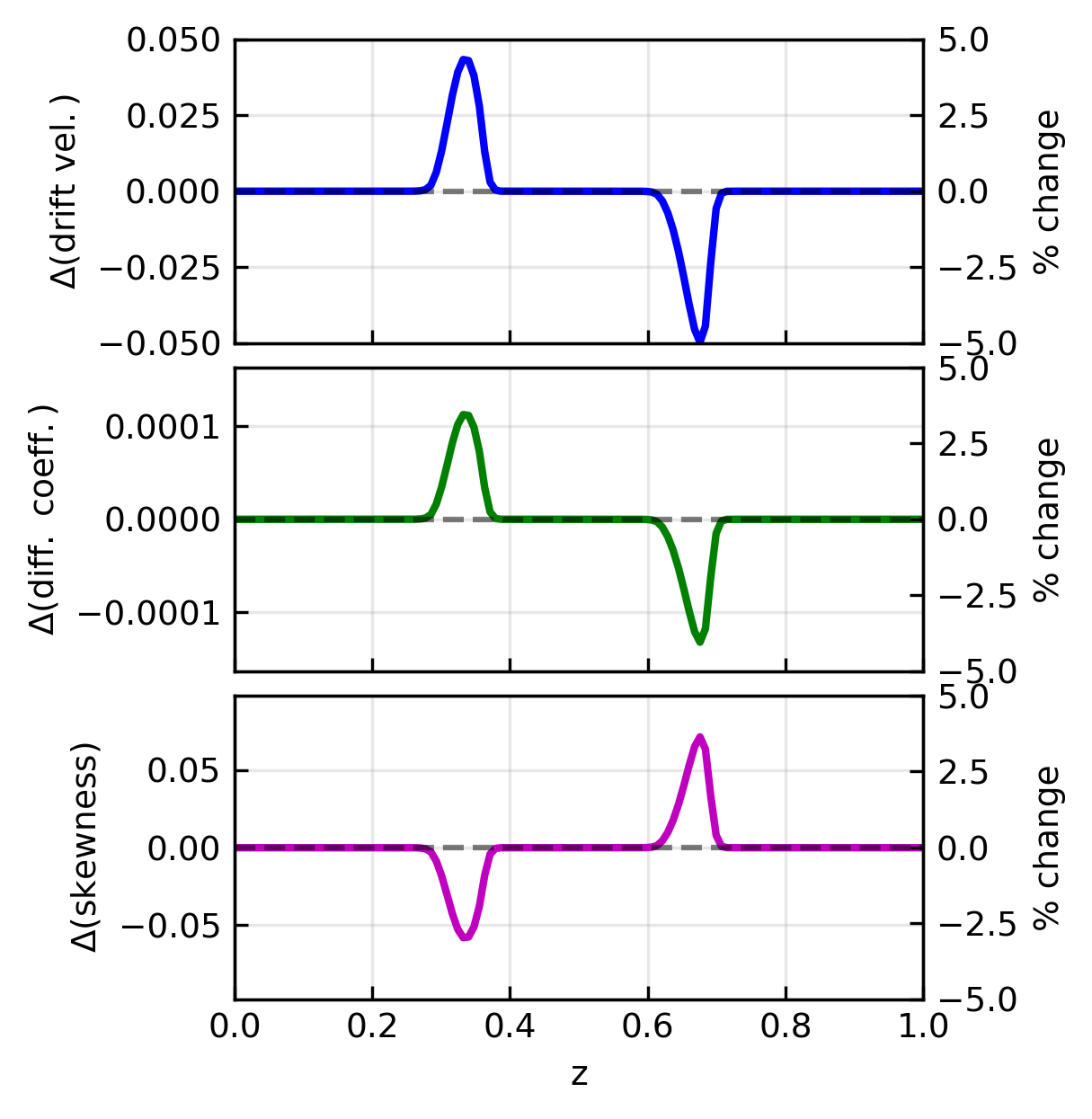}
    \caption{Changes in tracer velocity (top), diffusion coefficient (middle), and skewness of the kick probability density function (PDF; bottom) as a function of $z$ for our square pulse advection test (\S~\ref{sec:square_pulse}) with the second-order scheme with MonCen slope limiter. We compute these values directly using the formulas in Eqs.~\ref{eq:drift}-\ref{eq:skewness}. Small variations in velocity and diffusion coefficient are critical for getting the tracer particle density to match the gas density. Without these small variations at the pulse boundaries, we would have a constant advection velocity and diffusion coefficient, and the resulting tracer distribution would be identical to that in  Fig.~\ref{fig:square_pulse0}; that is to say, much \textit{too} diffusive. Allowing for these variations is critical to getting the correct tracer behavior.}
    \label{fig:square_pulse_deviations}
\end{figure}

Examining Fig.~\ref{fig:lagrangian_fidelity}, we see that all of the stochastic methods agree quite closely. They also align quite closely with the analytic solution, only differing visibly near the tails. The reason for this is that the analytic solution uses a constant, first order diffusion coefficient, while in the actual test, the advection velocity and diffusion coefficient the particles see varies over the transition regions from high to low density (or vice versa, Fig.~\ref{fig:square_pulse_deviations}). Over the single advection period we use here, the particles spread enough that a minority of them make it to this transition region.


\subsubsection{Statistical comparison of tracer methods}\label{sec:stat_comparison}

As all of our tracer methods (with the exception of the classical tracer) are stochastic in nature, their distributions cannot be rigorously compared to one another directly. Instead, we must perform statistical tests to determine how well they agree with one another. We present several such tests here. The first test relies on an assumption of Gaussianity: we simply compute the standard error of the standard deviation ${\rm SE}(\sigma)$ of the particle positions (initially located in a sheet, Sec.~\ref{sec:sheet}) and perform a standard $Z$-test using the resulting $Z$-scores.

The second test is a $\chi^2$ homogeneity test. It requires computing a $\chi^2$ statistic summed across the $128$ cell-centered sheets along $z$, each of thickness $h$. 

\paragraph{Pairwise Z-tests.}\label{par:z_tests} The $Z$-statistic, given two particle types $a, b$ is computed as,
\begin{align}
    Z_{ab} &\equiv \frac{\sigma_a - \sigma_b}{\sqrt{{\rm SE}(\sigma_a)^2 + {\rm SE}(\sigma_b)^2}},\\
    \sigma_a^2 &\equiv \sum_k \frac{(z_k-\bar{z}_a)^2}{N-1}\\
    \bar{z}_a &\equiv \sum_k\frac{z_k}{N},\\
    {\rm SE}(\sigma_a)&\equiv \frac{\sigma_a}{\sqrt{2(N-1)}},
\end{align}
where the sums are performed over all particles $k$ initially within the thin sheet, and $N =524,288$ is the number of particles (identical) in each sample. The pairwise results of this are shown in Tab.~\ref{tab:z_test}. The standard error for all particle types is approximately ${\rm SE} \approx 7.73\times 10^{-5}$. All of the $Z$-scores are $|Z| < 1$, implying that the distributions are statistically indistinguishable with this test. 

\begin{table}
  \centering
  \caption{Pairwise $Z$-tests on the various tracer types, initially located in a thin sheet, after advecting for one period in the square pulse test (Secs.~\ref{sec:sheet},~\ref{par:z_tests}). $\Delta \sigma$ is the difference between the standard deviation in $z$-position for the first particle type minus the second, E.g. for the pair $A$ vs $B$, $\Delta\sigma = \sigma_A - \sigma_B$. All $p$-values are two-tailed.}
  \label{tab:z_test}
  \begin{tabular}{lccc}
    \toprule
    Pair & $\Delta\sigma$ & $Z$ & $p$  \\
    \midrule
    MC vs It\^o-2
      & $+1.05\times10^{-4}$ & $0.96$ & $0.338$ \\
    MC vs It\^o-3
      & $+5.42\times10^{-5}$ & $0.50$ & $0.620$ \\
    It\^o-2 vs It\^o-3
      & $-5.06\times10^{-5}$ & $-0.46$ & $0.644$ \\
    \bottomrule
  \end{tabular}
\end{table}


\paragraph{$\chi^2$ homogeneity test.}
Assume, as a null hypothesis, that the different particle types are drawn from the same distribution. Then the density of particles in each cell should be consistent with one another, up to Poisson fluctuations. Under this assumption, we construct a $\chi^2$ test as follows. We effectively use $128$ bins for the particles, one per cell in the $z$-direction. The CIC-deposited particle count for $z$-index $k \in [1,128]$ and tracer scheme $a$ is $N_{a,k}$. If the null hypothesis is true, then we may pool the fractions of particles in each $z$-bin,
\begin{align}
    f_{k} &\equiv \frac{N_{a,k}+ N_{b,k}}{N_a + N_b},\\
    N_a &\equiv \sum_k N_{a,k}.
\end{align}
Then the expected number of particles in each bin $k$ for tracer type $a$ is,
\begin{align}
    E_{a,k} \equiv N_a f_k.
\end{align}
Given this expected count, we may construct a test statistic summed across all bins:
\begin{align}
    \chi^2 &\equiv \sum_k \left\{\frac{(N_{a,k}-E_{a,k})^2}{E_{a,k}} + \frac{(N_{b,k}-E_{b,k})^2}{E_{b,k}}\right\}.
\end{align}
Under the null hypothesis and assuming Poisson noise as the cause of the fluctuations, this should follow a $\chi^2$ distribution with degrees of freedom ${\rm df} = 128-1 = 127$. This test is much more sensitive than our $Z$-test, as it can detect differences in shape per bin. This test can be reasonably used for both the spread of the particle sheet initially located between $z = 63h$ and $64h$, as well as for the overall distribution of particles in the square pulse test (Sec.~\ref{sec:square_pulse}). The result of this test is shown in Tab.~\ref{tab:chi_squared}. 

When looking at only the particles initially within a sheet, we find that the MC and It\^o-3 tracer distributions are consistent with one another (per-degree-of-freedom $\chi^2/{\rm df} = 0.84$, $p = 0.88$), while the other tracer distributions are highly inconsistent with one another, generating $\chi^2/{\rm df} > 3$ and $p \lesssim 10^{-21}$. This illustrates that matching the third statistical moment of the MC tracer process increases consistency between the methods.

Generalizing to the entire set of particles in the square pulse test largely tells the same story, with $\chi^2/{\rm df} > 6$ when comparing MC vs It\^o-2 and It\^o-2 vs It\^o-3 tracers to one another. This generates $p \lesssim 10^{-96}$. Comparing MC vs It\^o-3, we obtain $\chi^2/{\rm df} \approx 1.5$ and $p \lesssim 10^{-4}$, showing that while this test finds these methods inconsistent with one another, they are much more similar to one another than they are to the It\^o-2 method. Again, matching the third moment of the MC jump brings the It\^o tracer into better agreement with the MC tracer. 

\begin{table}
  \centering
  \caption{Pairwise $\chi^2$ homogeneity tests across $128$ z-bins
           ($\mathrm{df} = 127$) for both the sheet (top) and the entire set of particles for the square pulse test (Sec.~\ref{sec:square_pulse}). 
           A reduced statistic $\chi^2/\mathrm{df} \approx 1$ indicates
           consistency with a common distribution.}
  \label{tab:chi_squared}
  \begin{tabular}{lcc}
    \toprule
    \multicolumn{3}{l}{\textit{Homogeneity test for sheet}}\\
    \midrule
    Pair & $\chi^2/\mathrm{df}$ & $p$ \\
    \midrule
    MC vs It\^o-2
       & $3.30$ & $2.3\times10^{-25}$ \\
    MC vs It\^o-3
        & $0.84$ & $0.88$ \\
    It\^o-2 vs It\^o-3
       & $3.03$ & $1.5\times10^{-21}$\\
    \midrule
    \multicolumn{3}{l}{\textit{Homogeneity test for all particles}}\\
    \midrule
    Pair  & $\chi^2/\mathrm{df}$ & $p$ \\
    \midrule
    MC vs It\^o-2 & 6.236 & $1.12\times 10^{-96}$ \\
    MC vs It\^o-3 & 1.526 & $1.23\times 10^{-4}$ \\
    It\^o-2 vs It\^o-3 & 6.707 & $1.12 \times 10^{-107}$\\
    \bottomrule
  \end{tabular}
\end{table}

\subsection{Test: Isothermal decaying turbulence}\label{sec:turb}

\begin{figure*}
    \centering
    \includegraphics[width=1.0\textwidth]{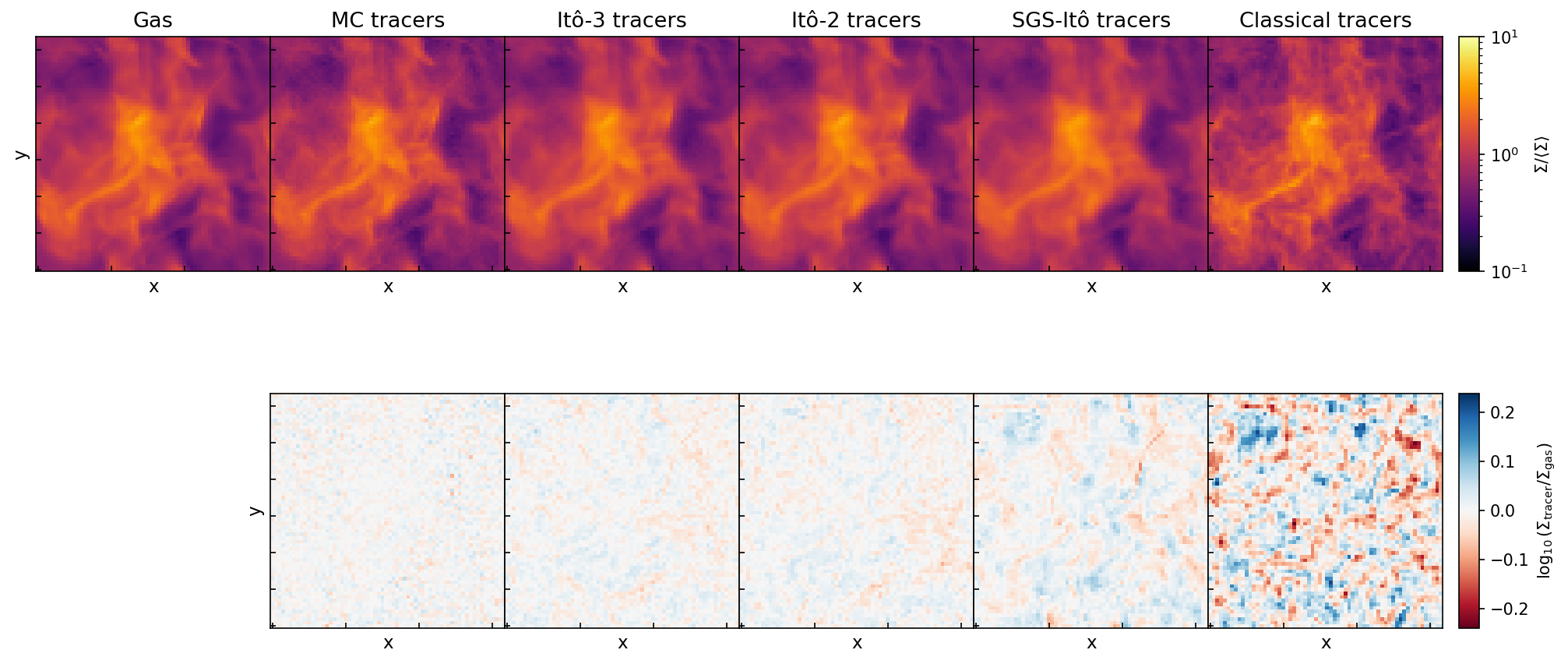}
    \caption{{\bf Top:} Column densities (normalized by the mean) of gas, MC, It\^o-3, It\^o-2, SGS-It\^o, and classical tracers for the decaying turbulence test described in Sec.~\ref{sec:turb}, taken at $t=0.2\,L_0/c_{\rm s}$. {\bf Bottom:} ratio of tracer column density to gas column density for the same particles. The MC tracers are those implemented and run in \textsc{RAMSES} \citep{cadiou2019accurate,teyssier2002cosmological}, while the other tracers here are run with identical initial conditions, but in the code \textsc{mini-RAMSES}.}
    \label{fig:turb_column}
\end{figure*}

\begin{figure*}
    \centering
    \includegraphics[width=1.0\textwidth]{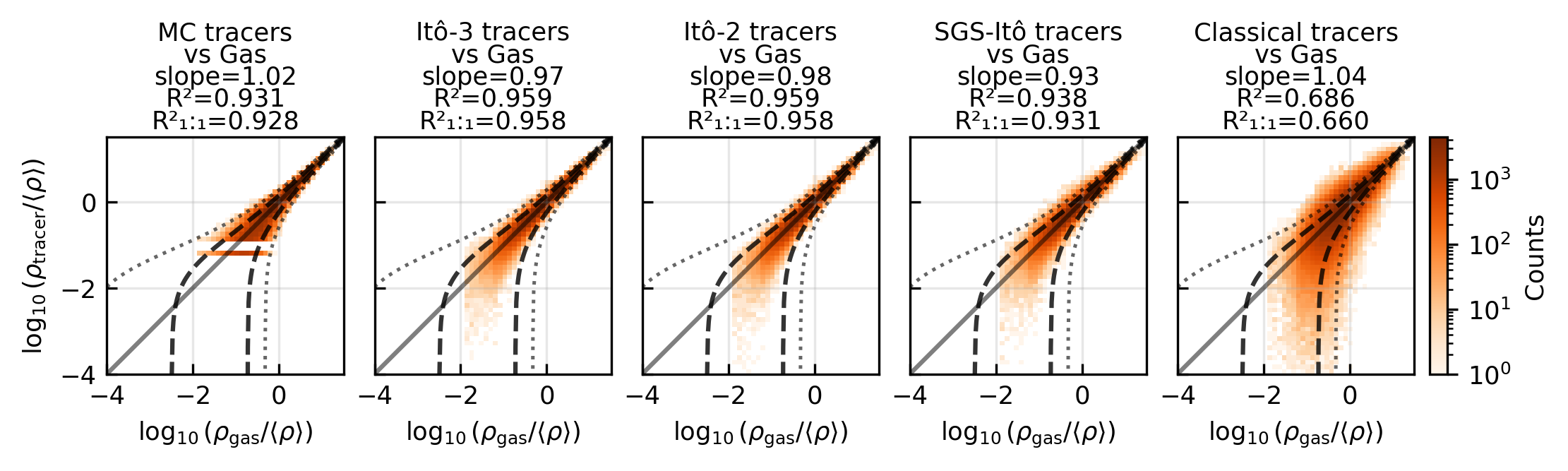}
    \caption{Joint density histograms of MC, It\^o-3, It\^o-2, SGS-It\^o, and classical tracers vs. gas. The solid gray line shows the 1:1 tracer:gas density. Dashed lines show the 90\% confidence region as computed from the Poisson distribution, while dotted lines show the 99.9\% confidence region. Each plot is also labeled with a best fit slope, an $R^2$ value for that slope, and an $R_{1:1}^2$ value computed assuming that the correct slope is 1 (cf. Sec.~\ref{sec:turb}).}
    \label{fig:joint_density}
\end{figure*}

Our second test is of decaying, 3D isothermal turbulence. We generate an initially solenoidal velocity field with a 3D r.m.s. equal to $\sigma_{\rm rms} = 15\,c_{\rm s}.$ The initial pressure and density are uniform, with $P_0 = \rho_0 = 1$, respectively. This also implies that the sound speed $c_{\rm s} = 1$. The initial power spectrum follows $k^{-5/3}$, with initial power running from $k_{\rm min} = 2\pi/L_0$ up to $k_{\rm max} = 5\times2\pi/L_0$, with zero power outside this range, where $L_0$ is the size of the box. We run the simulation until $t = 0.2\,L_0/c_{\rm s}$, i.e. for approximately $3$ turbulent turnover times. We use the HLLC Riemann solver and the MonCen slope limiter. The resolution of the simulation is $N_x = 64$ grid cells along each direction. We run the simulation both with \textsc{RAMSES} \citep{teyssier2002cosmological} and with \textsc{mini-RAMSES}\footnote{\nolinkurl{https://bitbucket.org/rteyssie/mini-ramses}}, resulting in identical output for the gas. We use \textsc{RAMSES} in order to use its implementation of MC tracers \citep{cadiou2019accurate}. As \textsc{RAMSES} and \textsc{mini-RAMSES} produce identical output, we may then directly compare our It\^o tracers as implemented in \textsc{mini-RAMSES} to the MC tracers in \textsc{RAMSES}. For all tracer schemes, we use 16 particles per cell. For the classical tracers, as they are deterministic, in order to have a higher particle count per cell, particles must be systematically displaced initially. \footnote{In order to maintain uniformity, we divide these tracers into 16 families, and compute 16 random offsets $\zeta_i \in [-1/2,+1/2]$, such that the particle positions are shifted by $\zeta_i h$, for $h = L_0/N_x$, with $L_0$ the box length and $N_x = 64$ the resolution. This essentially generates 16 superposed particle grids on top of one another, which when CIC deposited, will generate a uniform density.} For the SGS turbulence model (essential for the SGS-It\^o tracer, Sec.~\ref{sec:sgs_imc}), we choose the Smagorinsky-Lilly constant $C_{\rm s} = 0.17$, Lilly's theoretical value for isotropic, homogeneous turbulence \citep{lilly1966representation}.

\subsubsection{Tracer density: correlation and fluctuation}

\begin{figure}
    \centering
    \includegraphics[width=1.0\columnwidth]{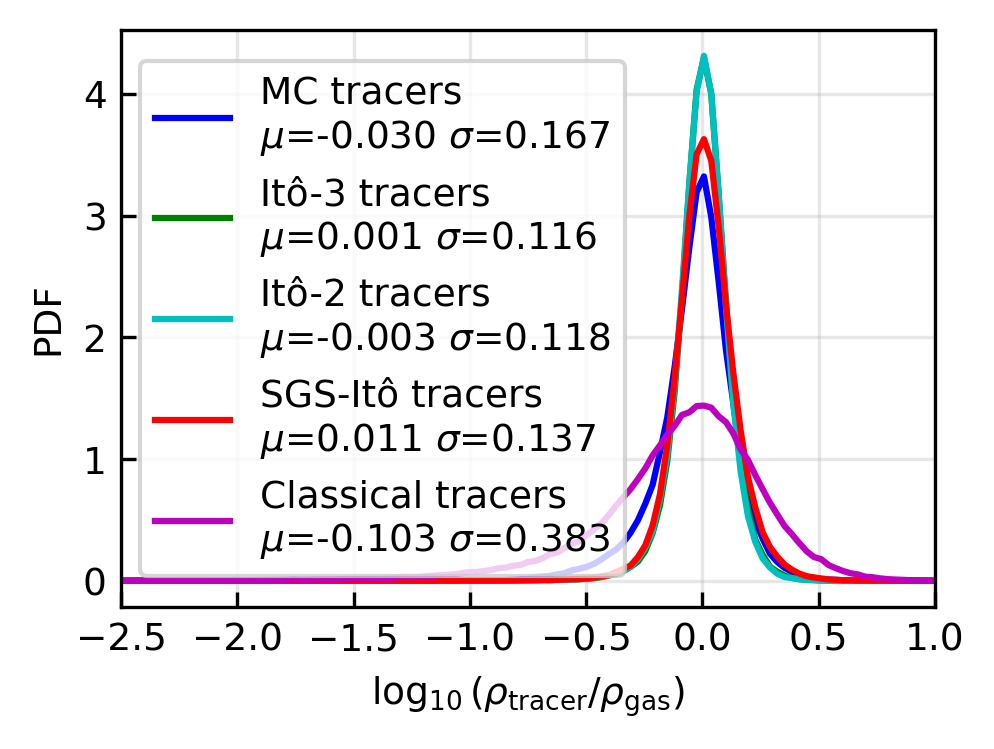}
    \caption{Probability density functions (PDFs) of the base-10 logarithm of the density of tracer particles relative to gas for our various methods (see legend). For each, we compute the mean of the PDF $\mu$, as well as its standard deviation $\sigma$. The It\^o-3 tracer shows the narrowest distribution, with the It\^o-2 tracer not far behind. Interestingly, the SGS-It\^o tracer is approximately on par with the MC tracer. The classical tracer is by far the worst performing, and also has a mean most discrepant with zero.}
    \label{fig:log_ratio}
\end{figure}

In Figure~\ref{fig:turb_column}, we show the column density of gas and the various tracer schemes (top row), and the column density of tracers relative to gas (bottom row). While the similarity between the various stochastic tracer schemes is evident, the accuracy hierarchy is also apparent. By eye, the various flavors of It\^o tracer appear to have their noise (bottom row) more spatially correlated, with It\^o-3 faring better than It\^o-2, It\^o-2 better than SGS-It\^o, and SGS-It\^o better than classical tracers. This illustrates that most of the gain is achieved through going from a deterministic (classical) tracer to a stochastic one (SGS-It\^o), with further improvements as one matches the gas numerical diffusion (It\^o-2) and a lesser gain in matching the numerical dispersion (It\^o-3). 

The tracer method that stands out the most is the classical tracer, which has no diffusion terms whatsoever. This means that as small-scale power accumulates, it has no means of dissipating. While the other tracer schemes still exhibit some numerical bias, such bias is non-persistent, as it is allowed to dissipate, a process which is impossible for the classical tracers.

We examine the density of tracers relative to gas more quantitatively in Figs.~\ref{fig:joint_density} and \ref{fig:log_ratio}. Fig.~\ref{fig:joint_density} shows the joint density histograms of ($\log_{10}$) tracer density versus gas density for MC, It\^o-3, It\^o-2, SGS-It\^o, and classical tracers. At the top of each panel, we also show the results of a (density-weighted) linear regression, in particular showing the slope as well as Pearson's correlation coefficient $R^2$, and also a value $R^2_{1:1}$ which is computed assuming the slope of the line is equal to 1. $R^2_{1:1}$ can thus be thought of as measuring how good of a fit the tracer data is to the gas data. This figure also shows a 90\% (dashed) and 99.9\% (dotted) confidence regions computed assuming tracer density is subject to Poisson noise, the magnitude of which is related to the gas density. The shape of this Poisson noise envelope matches the shape of the tracer PDFs quite well. It also gives us an objective basis of comparison between the various joint PDFs. For example, it can be seen that, while the noise is quite close to the Poisson model for the MC, It\^o-2, and It\^o-3 tracers, the SGS-It\^o tracer shows more deviation. In a sense, this method lies somewhere between the classical tracer and the It\^o tracers, both conceptually and in terms of performance.

All slopes for all tracer methods are $\approx 1$. It\^o tracers show a similar $R^2_{1:1} \approx 0.96$, regardless of the kick PDF shape. The MC tracers show $R^2_{1:1} \approx 0.93$, while the SGS-It\^o tracers show $R^2_{1:1} \approx 0.9$. Only the classical tracer shows a meaningfully worse $R^2_{1:1} \approx 0.66$. The reasonably good performance of the SGS-It\^o tracer, despite the fact that its diffusion comes from a completely different (and also, isotropic) model suggests that it matters less what exactly the diffusion looks like and more that it is present in some form. The SGS model will also tend to produce diffusion roughly where it is needed, and neglect to produce it where it is not: quiescent regions will have a low diffusion coefficient, while highly turbulent, compressed or shocked regions will have a high diffusion coefficient. 

In Fig.~\ref{fig:log_ratio}, we look at the logarithm of the tracer density relative to the gas density for the different tracer methods. We also list (in the legend) the mean of each distribution $\mu$ and the standard deviation $\sigma$. These values are calculated on the log-transformed values. As usual, the classical tracer has the worst performance, with the largest $|\mu|$ and $\sigma$. The narrowest PDF is the It\^o-3 tracer, followed closely by the It\^o-2 tracer; evidently, for this turbulence test, matching the third moment is not especially important. This makes sense, as the leading order error is corrected by adding in the appropriate diffusion alone, and the skewness is a higher order correction (cf. Fig.~\ref{fig:square_pulse}). The SGS-It\^o tracer has a similar PDF width to the MC tracer.

These findings are interesting because, as can be seen in Figs.~\ref{fig:square_pulse} and \ref{fig:turb_column}, the It\^o tracers remain slightly biased relative to the MC tracers in certain regions. Thus, despite being somewhat biased, they still have a better correlation. Furthermore, because the tracers necessarily are capable of dissipating structures through diffusion, such bias is \textit{non-persistent}, and thus can remain small in amplitude, in contrast to the classical tracers. This is one of the key reasons why, despite using the same cell-centered advection velocity as classical tracers, SGS-It\^o tracers still manage to improve vastly over them across all measures.

\subsubsection{Comparison of density powerspectra}

\begin{figure}
    \centering
    \includegraphics[width=1.0\columnwidth]{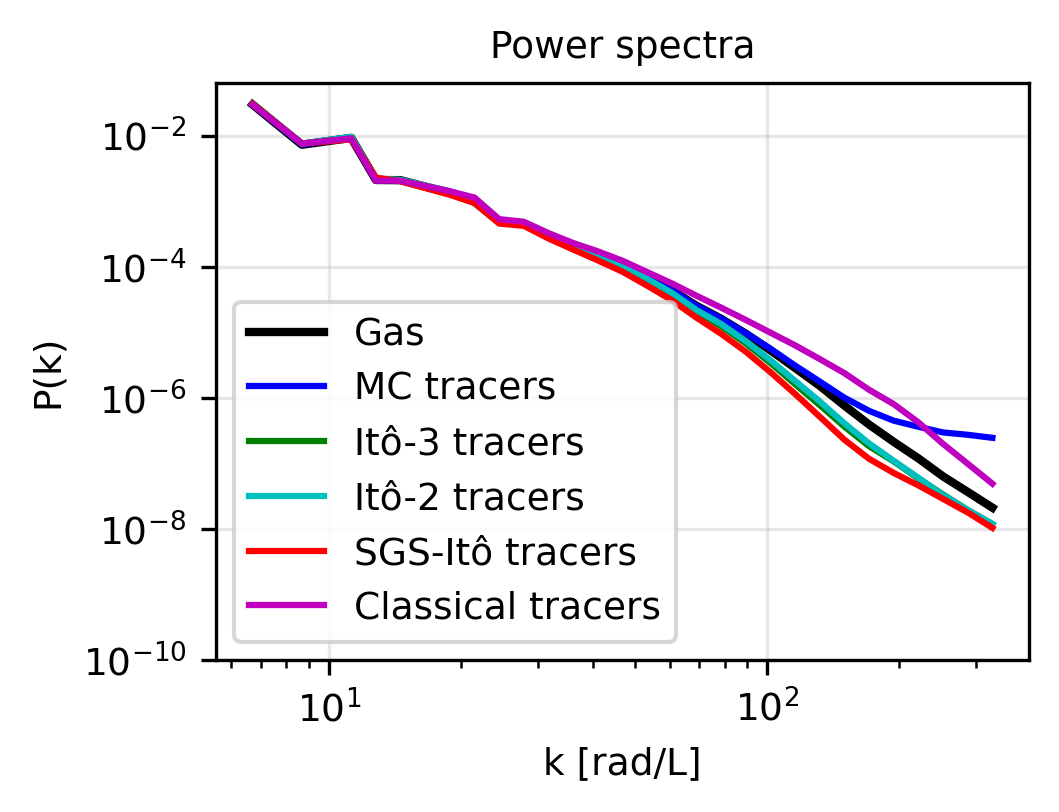}
    \caption{Density power spectra vs. wavenumber $k$ for the gas, as well as the various tracer methods. Classical tracers show elevated power across a large range of scales, while It\^o tracers tend to show suppressed power. MC tracers tend to have exact agreement in power until one reaches small scales, where Poisson noise creates increased power. On sufficiently large scales, all methods tend to agree with the gas.}
    \label{fig:powerspectrum}
\end{figure}

\begin{figure}
    \centering
    \includegraphics[width=1.0\columnwidth]{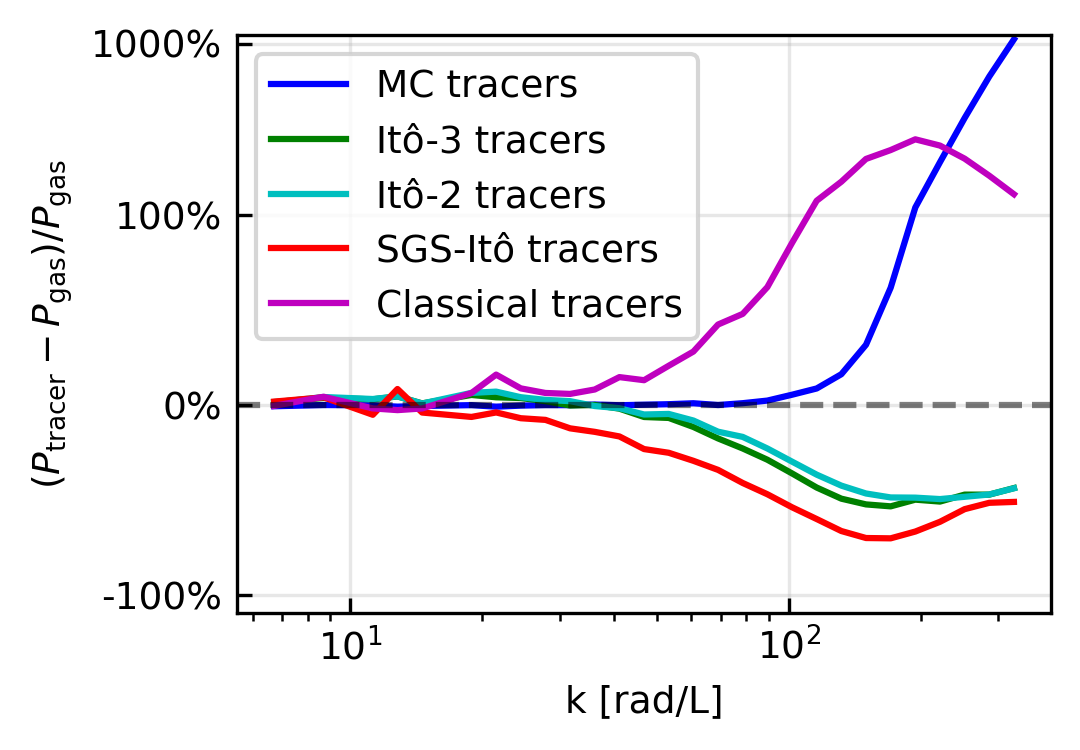}
    \caption{Similar to Fig.~\ref{fig:powerspectrum}, but highlighting the relative difference in the powerspectra relative to the gas power.}
    \label{fig:ps_rel_error}
\end{figure}

We compute density power spectra for the gas and various tracers, and show the results  in Fig.~\ref{fig:powerspectrum}, binned by wavenumber. To construct these spectra, we deposit the gas and tracer densities onto a uniform $N_x^3$ cube. The tracers are deposited with the same CIC kernel used elsewhere in this paper. From each field we form the overdensity $\delta = \rho/\langle\rho\rangle - 1$, take its three-dimensional discrete Fourier transform $\delta_{\mathbf k}$, and average $|\delta_{\mathbf k}|^2$ over logarithmically spaced spherical shells of $k = |{\mathbf k}|$ to obtain $P(k)$.

To highlight the differences between the methods, we have also computed the relative difference between these powerspectra in Fig.~\ref{fig:ps_rel_error}. The various It\^o methods all show less power than the gas as one approaches smaller scales, implying they are actually somewhat \textit{more} diffusive than the gas, and overly smooth the small-scale structure. We believe this to at least be partly due to the CIC kernel deposition smoothing out cell-scale structures in the particle distribution. On the other hand, Poisson noise dominates the small scales for the MC tracer, meaning the small-scale power in the tracers is actually greater than that for the gas; however, as is shown in \citet{cadiou2019accurate}, this additional small-scale noise is not biased.

On the other hand, classical tracers show excess power across virtually all length scales in the simulation, and this power is known to be non-random, and systematic \citep{genel2013following,cadiou2019accurate}. Consistent with \citet{price2010comparison} and in light of our other findings in this paper, we believe this phenomenon is essentially due to the fact that classical tracers have no diffusion, which causes an accumulation of small-scale power.


\section{Summary \& Conclusions}\label{sec:summary}
In this paper, we have presented a continuous-time Lagrangian tracer particle method, which we call the It\^o tracer. We began (Sec.~\ref{sec:ito1d}) by showing how the distribution of an ensemble of particles maps onto the evolution of a fluid, and how, in order to match the fluid's numerical diffusion, a random kick of the appropriate size must be included at each step. The reason classical tracers fail to reproduce the density of the fluid is this: the fluid experiences numerical diffusion, and the classical tracer has none. A classical tracer is an indivisible point; it cannot spread out in space, and so it cannot match the density field of the fluid. A stochastic particle can, albeit only in a statistical sense.
We determined the magnitude that this diffusion must have by matching the first two moments of the Markov transition kernel of the numerical hydrodynamics scheme, as realized discretely by the Monte-Carlo (MC) tracer jump of \citet{genel2013following} and \citet{cadiou2019accurate}: the first moment sets a deterministic drift, and the second sets the strength of a stochastic L\'evy kick. We refer to this two-moment-matched variant as the It\^o-2 tracer. Its drift is built from the Riemann solver's mass fluxes (Eq.~\ref{eq:drift}) rather than from the cell-centered gas velocity, and this is what cures the density-concentration failure of classical velocity tracers. In the incompressible limit, the corresponding diffusion coefficient reduces to the first-order upwind numerical diffusion of the underlying Godunov scheme (Eq.~\ref{eq:fo_upwind_diff}). 

We extended the moment match to the third moment of the transition kernel by introducing a piecewise skew uniform (PSU) kick PDF (Sec.~\ref{sec:ito3}, Fig.~\ref{fig:psu}), yielding the It\^o-3 tracer, which brings the method into closer statistical agreement with the MC tracer, but at the cost of losing formally continuous paths. By Pawula's theorem \citep{pawula1967approximation,risken1989fokker}, a continuous-path Markov process cannot match a nonzero third Kramers--Moyal coefficient, so It\^o-$n\ge3$ is a c\`adl\`ag jump-diffusion (Sec.~\ref{sec:iton_continuity}). These methods occupy a single family of processes: It\^o-2 at the pure-diffusion end, MC at the pure-jump end, and It\^o-$n\ge3$ tracers with nonzero higher-order coefficients as jump-diffusion hybrids.
As discussed in Sec.~\ref{sec:intro}, this moment-matching construction places the It\^o tracer within the broader Lagrangian-SDE tradition developed for turbulent reactive flows \citep{lundgren1967distribution,dopazo1974approach,pope1985pdf,haworth2010progress}, but with a discrete, bounded, skewed target process rather than a closure-modeled turbulent flow.

In multiple dimensions, the fact that the MC face jumps are mutually exclusive produces negative cross-correlation at finite $\delta t$, even though the raw second moment tensor is purely diagonal. Appendix~\ref{app:finite_covariance} derives these terms and shows that their impact is minute in practice. 

The subgrid-scale It\^o tracer (SGS-It\^o; Sec.~\ref{sec:sgs_imc}) shows that we can achieve comparable performance with a diffusion that is physical in origin rather than numerical \citep{smagorinsky1963general,lilly1966representation,wadsley2008treatment}. This is true even though the SGS-It\^o particle is advected by a CIC-interpolated cell-centered velocity, rather than by the downwind flux-based velocity used by the It\^o tracer. The lesson is that, to leading order, the specific form of the advection and diffusion matters less than whether any diffusion is present at all. Without it, tracer particles form persistent, arbitrarily sharp features \citep{price2010comparison, genel2013following,cadiou2019accurate}. With it, those features necessarily relax over time, or else are prevented from forming in the first place.

It is worth stressing that the SGS-It\^o tracer uses the same cell-centered velocities that the classical tracer does. Though we have opted not to show the results in this paper, we tested a variant of the classical tracer that uses the same advection velocity as the It\^o tracer (Eq.~\ref{eq:drift}), but with zero stochastic kick/diffusion, to test the hypothesis that the advection is the main source of mismatch between gas and tracers. This not only did not improve the mismatch: it actively made it worse. Thus, this velocity is only the correct velocity to use in the context of the overall It\^o scheme.

We tested the It\^o and SGS-It\^o tracers against the \textsc{RAMSES} MC tracer implementation in two settings. In a 1D square-pulse advection test (Sec.~\ref{sec:square_pulse}), pairwise $Z$-tests on the width $\sigma$ of the advected sheet cannot distinguish any of the stochastic methods from one another (Tab.~\ref{tab:z_test}; all $|Z| < 1$). A pairwise $\chi^2$ homogeneity test across the $128$ $z$-bins, sensitive to the shape of the distribution rather than its width alone, accepts the common-distribution null hypothesis only for MC vs.\ It\^o-3 ($\chi^2/\mathrm{df} = 0.84$, $p = 0.88$; $\mathrm{df} = 127$); pairs involving the It\^o-2 tracer are rejected ($\chi^2/\mathrm{df} \approx 3$, $p \lesssim 10^{-21}$). The same ordering is recovered when the test is applied to the full pulse rather than to the advected sheet, with MC closer to It\^o-3 than to It\^o-2 by a factor of $\sim\!4$ in $\chi^2/\mathrm{df}$ (Tab.~\ref{tab:chi_squared}). In the more demanding 3D $\sigma_{\rm rms} = 15\,c_{\rm s}$ decaying isothermal turbulence test (Sec.~\ref{sec:turb}), where the gas is evolved in both \textsc{RAMSES} \citep{teyssier2002cosmological} and \textsc{mini-RAMSES} to identical output, the It\^o tracers reproduce or improve upon the MC tracer statistics with the same number of particles per cell (Figs.~\ref{fig:turb_column},~\ref{fig:joint_density},~\ref{fig:log_ratio},~\ref{fig:powerspectrum}). We should note, however, that as one needs to compute the same quantities as the MC tracer, but in each of the eight cells the CIC kernel overlaps with, the It\^o tracer is more expensive than the MC tracer. It is also slightly more expensive than the classical tracer, as we need to draw one random number per particle at each step in order to evolve them.
A large part of the value of the It\^o tracer is methodological. Unlike the classical tracer or the MC tracer, it provides a \textit{framework} into which a wide variety of numerical and physical processes may be incorporated. Because the particles follow an SDE, the method maps cleanly onto other continuous-trajectory particle types --- dust grains, charged particles, cosmic rays --- for which a discrete-jump scheme has no obvious counterpart.

The SDE framing also opens a number of future directions that are inaccessible to discrete-jump MC schemes. The associated Fokker--Planck equation can be solved analytically for simple reference problems, providing exact benchmarks against which particle histograms may be compared (cf. Eq.~\ref{eq:green_function} and Figs.~\ref{fig:square_pulse0},~\ref{fig:square_pulse}). The It\^o method is compatible with standard variance-reduction techniques, such as antithetic variates, common random numbers, Brownian-bridge path construction, quasi-Monte-Carlo seeding, and higher order integration methods (e.g. Milstein, Sec.~\ref{sec:algorithm}). None of these techniques apply straightforwardly to a discrete-jump process. It is differentiable with respect to the underlying flow state, which enables gradient-based sensitivity analysis and machine-learning-based subgrid closures. Finally, the continuous, branch-free particle update is more amenable to GPU execution than the data-dependent cell-exchange structure of MC tracers.
To summarize:
\begin{enumerate}
    \item[(i)] We introduced a continuous-time Lagrangian tracer by matching an It\^o stochastic differential equation (SDE) to the Markov transition kernel of the numerical hydrodynamics scheme, as realized by the discrete MC jump of \citet{genel2013following} and \citet{cadiou2019accurate} (Sec.~\ref{sec:theory}). It\^o-$n$ matches the first $n$ displacement moments of the 1D kernel. In multiple dimensions, the construction is applied independently along each coordinate and therefore matches the marginal, rather than complete finite-step joint, moments (Appendix~\ref{app:finite_covariance}). Only It\^o-2 has almost-surely continuous paths; every It\^o-$n$ process with $n\ge3$ and a nonzero higher-order coefficient is a c\`adl\`ag jump-diffusion (Sec.~\ref{sec:iton_continuity}).
    \item[(ii)] The It\^o drift is derived from the Riemann solver's mass fluxes (Eq.~\ref{eq:drift}) rather than from the cell-centered gas velocity, curing the density-concentration failure of classical velocity tracers. Its diffusion coefficient follows from the MC-jump variance (Eq.~\ref{eq:diffusion}).
    \item[(iii)] For a second-order numerical scheme (using the MonCen slope of the \textsc{RAMSES} code), the It\^o-3 tracer, which matches the third moment of the MC jump via a piecewise skew uniform (PSU) kick distribution (Sec.~\ref{sec:ito3}, Fig.~\ref{fig:psu}), is in statistical agreement with the MC reference on a pairwise $\chi^2$ homogeneity test across the advected sheet ($\chi^2/\mathrm{df} = 0.84$, $p = 0.88$; $\mathrm{df} = 127$), while the It\^o-2 tracer is rejected by the same test ($\chi^2/\mathrm{df} \approx 3$, $p \lesssim 10^{-21}$), despite a width $\sigma$ indistinguishable from the MC tracer at the $Z$-test level (Tab.~\ref{tab:z_test} and~\ref{tab:chi_squared}).
    \item[(iv)] The SGS-It\^o variant (Sec.~\ref{sec:sgs_imc}), based on subgrid-scale isotropic turbulent diffusion \citep{smagorinsky1963general,lilly1966representation,wadsley2008treatment} rather than on numerical diffusion, achieves performance comparable to the MC and It\^o schemes (Figs.~\ref{fig:joint_density},~\ref{fig:log_ratio}), reinforcing the broader point that the presence of diffusion in the first place matters more than its exact form.
    \item[(v)] We validated our It\^o tracer implementation against the \textsc{RAMSES} MC tracer implementation in a 1D square-pulse advection test and a 3D $\sigma_{\rm rms} = 15\,c_{\rm s}$ decaying isothermal turbulence simulation. In both cases it reproduces or improves upon the MC statistics (Tab.~\ref{tab:z_test},~\ref{tab:chi_squared}; Fig.~\ref{fig:square_pulse},~\ref{fig:lagrangian_fidelity},~\ref{fig:turb_column},~\ref{fig:joint_density},~\ref{fig:log_ratio},~\ref{fig:powerspectrum}).
    \item[(vi)] The SDE formulation places It\^o tracers within the broader Lagrangian-SDE framework developed for turbulent reactive flows \citep{lundgren1967distribution,dopazo1974approach,pope1985pdf,haworth2010progress}, and opens access to variance-reduction and higher order integration techniques that are unavailable to discrete-jump tracer methods.
\end{enumerate}

\section*{Data availability}
The data presented in this paper was generated on a MacBook Pro (14'', Nov 2023) and will be made freely available upon request to the corresponding author. Simulations with MC tracer particles were performed using the \textsc{RAMSES} code available on GitHub. Simulations with our new It\^o tracer particles were performed with the \textsc{mini-RAMSES} code available on BitBucket.

\section*{Acknowledgments}
We acknowledge helpful discussions with Susan Clark on this topic. We also acknowledge discussion with Shy Genel, whose feedback improved some of the figures, and Troels Haugbølle, further discussion with whom has refined our thinking on this topic. We thank Drummond Fielding for identifying the finite-step covariance distinction discussed in Appendix~\ref{app:finite_covariance}.
This work was supported by the U.S. Department of Energy SLAC
Contract No.DE-AC02-76SF00515. RT acknowledges support from the National Science Foundation (NSF) and the
U.S.-Israel Binational Science Foundation (BSF) under Award Number 2406558 and Award Title ``The Origin of the Excess of Bright
Galaxies at Cosmic Dawn''.

\bibliographystyle{mnras}
\bibliography{ms_reduced}

\appendix

\section{Finite-step covariance in multiple dimensions}
\label{app:finite_covariance}
In multiple dimensions, the It\^o-$n$ generalization presented in Sec.~\ref{sec:ito3d} is not complete at finite time-step. The reason for this is that mass transfer between cells only happens in one of six directions, and never in the diagonal directions. This distinction manifests as a difference between the raw second moment (Eq.~\ref{eq:raw2nd}) and the central covariance matrix. We derive the relevant terms and explore and bound their effects on the results of Sec.~\ref{sec:turb}.

{Let $h$ be the cell width, and let $C_{\pm,i}$ be the corresponding quantities in Eqs.~\ref{eq:cplus} and \ref{eq:cminus}. A step leaves the tracer in place with probability $1-\sum_iC_{+,i}$. Because one MC transition crosses at most one face, the mean displacement and raw second moment are}
\begin{align}
    m_i &\equiv \mathbb{E}[\Delta X_i]
    =hC_{-,i},\label{eq:app_mean}\\
    R_{ij} &\equiv \mathbb{E}[\Delta X_i\Delta X_j]
    =h^2C_{+,i}\delta_{ij}.\label{eq:app_raw}
\end{align}
The raw cross moment vanishes when $i\neq j$, while the central second moment remains finite:
\begin{align}
    Q_{ij}
    &\equiv
    \mathbb{E}\!\left[(\Delta X_i-m_i)(\Delta X_j-m_j)\right]\nonumber\\
    &=R_{ij}-m_im_j\nonumber\\
    &=h_i^2C_{+,i}\delta_{ij}
      -h_ih_jC_{-,i}C_{-,j}.\label{eq:app_covariance}
\end{align}
{The diagonal elements, $Q_{ii}=h_i^2(C_{+,i}-C_{-,i}^2)=2\kappa_i\Delta t$, recover the 1D variance in Eq.~\ref{eq:diffusion}. The off-diagonal entries, $-h^2C_{-,i}C_{-,j}$, are nonzero when the fluid is advected along more than one dimension. They arise because the face jumps of an MC tracer are mutually exclusive: a jump across an $x$ face precludes a jump across a $y$ face during the same step, so the two displacements are anti-correlated. For uniform diagonal flow with $C_x=C_y=C$, the correlation is}
\begin{align}
\operatorname{Corr}(\Delta X,\Delta Y)
    =-\frac{C}{1-C}.\label{eq:app_2d_corr}
\end{align}

{In the infinitesimal Kramers--Moyal limit, the distinction is moot. The expected displacement $m_i=u_i\Delta t=\mathcal{O}(\Delta t)$, thus}
\begin{align}
    \lim_{\Delta t\rightarrow 0}\frac{m_im_j}{2\Delta t}
    =\mathcal{O}(\Delta t)\longrightarrow0.
\end{align}
Therefore, the off-diagonal elements of the covariance matrix vanish in the limit of small time-step, which is why we adopt the diagonal form in Sec.~\ref{sec:ito3d}. Yet, in practice, these cross terms do have a finite (albeit small) impact on the method.

To match the central second moment, we would need to replace the diagonal diffusion of Eq.~\ref{eq:3d_kappa_diag} with $\kappa^{\rm step}_{ij}\equiv Q_{ij}/(2\Delta t)$, and draw $\Delta{\mathbf X}={\mathbf m}+{\mathbf L}{\bm\xi}$, where ${\mathbf L}{\mathbf L}^{\rm T}={\mathbf Q}$ and ${\bm\xi}$ is a vector with independent components that each have zero mean and unit variance. 

This correction to the method comes with a few caveats. First, it is more expensive than drawing purely independent kicks. It also changes higher order joint moments, which means that the It\^o-$n$ tracer matches marginal, rather than mixed, higher order moments. 

Drummond Fielding (private communication) independently implemented both the diagonal and full-covariance updates in the hydrodynamics code \textsc{AthenaK}. The correction changes the ensemble diagnostics of a comparable decaying-turbulence test by at most ${\sim}1\%$, leaving the conclusions of Sec.~\ref{sec:turb} unchanged. We therefore retain the diagonal diffusion method and describe it as matching the marginal moments of the MC displacement rather than its complete finite-step covariance.

\end{document}